\documentclass[pdflatex,sn-vancouver-num]{sn-jnl}

\usepackage[utf8]{inputenc}
\usepackage{graphicx}
\usepackage{booktabs}
\usepackage{multirow}
\usepackage{amsmath,amssymb,amsfonts}
\usepackage{fvextra}
\usepackage{tcolorbox}
\usepackage{url}
\usepackage{enumitem}
\usepackage{tabularx}
\usepackage{float}
\usepackage{placeins}
\usepackage{pdflscape}
\usepackage{longtable}
\usepackage{hyperref}
\setlist[itemize]{leftmargin=1.5em}

\DefineVerbatimEnvironment{codeblock}{Verbatim}{breaklines=true,breakanywhere=true,fontsize=\small}

\begin{document}

\title[ClinGuard]{ClinGuard: a governance framework for reliable, transparent, and auditable clinical coordination}

\author*[1]{\fnm{Kaiyuan} \sur{Wu}}\email{kw459@duke.edu}
\author[2]{\fnm{Aditya} \sur{Nagori}}
\author[3,4]{\fnm{Nicoleta J.} \sur{Economou}}
\author[5]{\fnm{Robert M.} \sur{Califf}}
\author*[2,6,7]{\fnm{Rishikesan} \sur{Kamaleswaran}}\email{r.kamaleswaran@duke.edu}

\affil[1]{\orgname{Department of Computational Biology and Bioinformatics, Duke University School of Medicine, Durham, NC, USA}}
\affil[2]{\orgname{Department of Surgery, Duke University School of Medicine, Durham, NC, USA}}
\affil[3]{\orgname{Duke Health AI Evaluation and Governance Program, Durham, NC, USA}}
\affil[4]{\orgname{Department of Biostatistics and Bioinformatics, Duke University School of Medicine, Durham, NC, USA}}
\affil[5]{\orgname{Department of Medicine, Duke University School of Medicine, Durham, NC, USA}}
\affil[6]{\orgname{Department of Electrical and Computer Engineering, Duke University, Durham, NC, USA}}
\affil[7]{\orgname{Department of Biomedical Engineering, Duke University, Durham, NC, USA}}

\abstract{ClinGuard is a model-agnostic governance framework that treats clinical coordination as patient-scoped, policy-bound transactions rather than free text. An independent sentinel layer validates and governs agent-proposed actions, escalates when needed, and executes approved actions through limited-privilege connectors. Each transaction is linked to a tamper-evident audit record, while an outcomes monitor informs future autonomy decisions. Using MIMIC-IV-FHIR and skilled-nursing-facility placement, we distinguish retrospective demonstration from proposed prospective evaluation in this work.}


\maketitle

\section{Introduction}
\label{sec:introduction}

Health systems are becoming increasingly digitized and data-rich, yet many everyday coordination tasks remain open-loop. Referrals are placed without confirmed receipt, test results lack timely follow-up, and discharge steps rely on manual work across inboxes, faxes, and portals \cite{ref1,ref2,ref3,ref5}. These gaps create safety and throughput risk across a family of coordination workflows. Discharge to a skilled nursing facility is one well-documented example: case managers assemble packets, reconcile payer constraints, and follow up acknowledgements across external systems, and placement delays become a rate-limiting step that prolongs length of stay \cite{ref6,ref7}. The same failure mode appears across the other workflows. In hospital-to-home transitions, missed follow-up and medication discrepancies drive avoidable readmissions; in specialty referral, patients are lost in the gap between primary care and specialist scheduling. In prior authorization and medication refill, physicians spend roughly fourteen hours weekly on paperwork that delays serious adverse-event response and laboratory follow-up and increases malpractice exposure \cite{ref8,ref9,ref10}. In short, many coordination failures are not knowledge deficits but missing state tracking, verification, and closure.

Large language models can draft notes and messages, but a prompt-to-text response is an unreliable proxy for work completion. Generative models are probabilistic and can hallucinate \cite{ref11,ref12,ref15,ref16}. In software supply chains, this vulnerability has facilitated slopsquatting, an attack class in which a language model invents a non-existent software-package name and an adversary later registers that name to deliver malware; the clinical analogue is a coordinator that cites a non-existent order set, guideline, or institutional policy that a downstream system might then act on as if it were real \cite{ref13}. Such a coordinator can also omit a prerequisite or imply a task is complete when no state change has occurred. A subtler failure arises when a copilot is optimized for a measurable proxy: it can learn to game that proxy, producing text that reads as a completed step without the underlying action, a pattern known as reward hacking. Generated text, like other health-IT outputs, can silently mask system-related errors \cite{ref14}. Operational reliability therefore requires a controllable execution system with an independent supervisory module that treats actions as verifiable state transitions, not as text alone.

In this light, we propose ClinGuard, a model-agnostic, externally auditable governance layer that turns clinical coordination into governed transactions: patient-scoped state transitions that are (i) grounded in auditable source data, (ii) validated against versioned institutional and payer constraints, (iii) executed only through limited-privilege connectors, which hold only the minimum permissions each action requires, and (iv) monitored to closure with escalation on delay or uncertainty. Where a coordinating agent is present (typically a language-model assistant), ClinGuard treats that agent's outputs as inputs to be validated, wrapping them in four sentinel functions before any state-changing execution. As outlined in Table~\ref{tab:3}, the supervisory boundary is sharp: the coordinating agent supplies semantics, ClinGuard's sentinel supplies the deterministic verification. This supervisory layer stays fixed even as the agent beneath it is upgraded, replaced, or retuned. That stability is what lets a deployment meet the formal change-control regimes it must answer to, the FDA Predetermined Change Control Plan \cite{ref34} and the NIST AI Risk Management Framework \cite{ref36}, as detailed in Section~\ref{sec:operationalization-in-clinics-integration-governance-and-aut}.

At a structural level, ClinGuard organizes governance around four sentinel functions: monitor, validate, govern, and escalate. How strictly each function acts is tuned to the workflow's risk. Section~\ref{subsec:four-sentinel-functions-and-risk-tier-aware-design} works through this tuning function by function.

This paper makes four contributions. First, a perspective shift from copilot-based assistance to governed coordination transactions with explicit state, closure criteria, and auditability (Table~\ref{tab:1}). Second, a clean separation between proposing and checking: whatever coordinating agent runs below ClinGuard drafts proposals, and the sentinel independently validates each one for policy compliance, factuality, and safety, judging it against a frozen data snapshot and versioned rules before anything executes. Third, a three-module supervisory architecture for procurement-grade deployment: (A) a pathway engine producing actions, (B) a ledger that produces an audit-grade decision record for every transaction with tamper-evident binding, and (C) an outcomes monitor that governs autonomy over time. Fourth, an ongoing pilot proposal at Duke University, with an evaluation design that measures the gains from governed execution over retrieval and drafting alone.

\paragraph{Scope of evidence.} ClinGuard is presented as a framework and proof of concept. Its architecture, contracts, and worked examples are implemented over MIMIC-IV-FHIR and exercised through an operator console (Supplementary Appendix~F), not in live care. Table~\ref{tab:evidence} separates capabilities demonstrated retrospectively, specified but not implemented, and reserved for prospective evaluation. We make no claim of prospective clinical efficacy, safety benefit, or workload reduction.

\section{Related work and positioning}
\label{sec:related-work-and-positioning}

Current artificial intelligence tools assist clinicians with documentation and decision-making but lack the capacity for independent action. Copilots draft notes, evidence assistants retrieve guidelines, and rule-based systems flag errors. However, these rules often fail to generalize across different hospitals, and none of these applications execute the actual coordination of care. Ensuring timely patient placement requires that referrals are transmitted, acknowledged, and tracked to closure.

The closest prior work involves \emph{Sentinel Agents}, which observe message exchanges between artificial intelligence agents to flag prompt injections, hallucinations, or data leaks. In one demonstration, they intercepted all 162 staged attacks, though the lack of benign case testing leaves the false-positive rate unknown \cite{ref33}. However, that approach only guards conversational exchanges. It does not govern concrete actions such as writing orders or sending referrals, which are the critical steps in a hospital setting. To address this, ClinGuard's sentinel checks each proposed action against the institution's current policies, records why it approved or blocked it, and adjusts how much the system is allowed to do on its own. Beyond medicine, general-purpose oversight architectures propose the same external, model-agnostic, real-time, immutable-ledger layer to supervise frontier models at societal scale (see also \cite{ref41}); ClinGuard is the clinical, FHIR-native, FDA- and NIST-aligned instantiation of that pattern.

ClinGuard is built as a separate supervisory layer. While a coordinating assistant handles drafting and summarizing, ClinGuard enforces reliability. Because this layer relies on deterministic rules, its behavior is highly reproducible across different clinical sites. ClinGuard does not attempt to provide clinical advice. Its primary value is strict enforcement at the point of action, backed by a complete audit trail and continuous outcome measurement that dictates the system's trusted level of autonomy. Table~\ref{tab:1} compares ClinGuard with single-agent copilots, Table~\ref{tab:2} contrasts it with the main families of governance and supervision approaches already in use, and Table~\ref{tab:3} shows how its sentinel relates to the agent it supervises.

The principal distinction is the enforcement point. Human review can catch unsafe suggestions but depends on attention, often leaves no structured rationale, and is vulnerable to automation bias \cite{ref51}. Clinical decision-support governance determines which rules are approved \cite{ref49} but does not adjudicate each firing. Model registries document systems before deployment \cite{ref50}, while machine-learning operations typically detects population-level drift after execution \cite{ref47}. ClinGuard instead evaluates each proposed state-changing action before execution, records a transaction-level decision, and can automatically reduce permitted autonomy.

\section{Design principles}
\label{sec:design-principles}

\subsection{Put the language model inside a controllable system}
\label{subsec:put-the-llm-inside-a-controllable-system-llm-neq-system}

Adopting separation of duties, we restrict the language model coordinator agent to tasks it performs well: semantic extraction, drafting, and candidate generation. Deterministic components within ClinGuard handle compliance, authorization, and execution. This structure enables testability and repeatable governance.

\subsection{Transactions, not chats}
\label{subsec:transactions-not-chats}

The fundamental unit of work in ClinGuard is a governed coordination transaction. This constitutes a state transition with explicit inputs, outputs, preconditions, and closure criteria. The system treats text as an explanation artifact rather than evidence of completion.

\subsection{Compliance-by-construction with versioned governance artifacts}
\label{subsec:compliance-by-construction-with-versioned-governance-artifac}

ClinGuard encodes institution and payer constraints as versioned policies and pathway clauses. Each executed transaction binds to these versions to enable reproducibility, incident review, and controlled change management. The system translates narrative guidelines and standard operating procedures into a typed normative intermediate representation with clause-level provenance. It then deterministically verifies these artifacts for consistency and conflict precedence before packaging them as semantically versioned outputs. Specific identifiers in code and schemas (PathwaySpec, Action Pack, Decision Dossier, TrustLedger, OutcomeSpec, EvidenceItem, ContextSnapshot) are referenced in the supplementary appendix.

\subsection{Least-privilege execution via limited-privilege connectors}
\label{subsec:least-privilege-execution-via-constrained-connectors}

Execution occurs strictly through privilege-scoped connectors authorized by ClinGuard. High-risk clinical orders remain draft-only unless explicitly signed by an authorized clinician.

\subsection{Auditability as a first-class output}
\label{subsec:auditability-as-a-first-class-output}

Every transaction produces an audit-grade record detailing the proposal, validation, execution, and timing. ClinGuard designs auditability as a primary product output combining a decision record and a tamper-evident audit ledger rather than reconstructing it post hoc.

\subsection{Outcomes close the loop and govern autonomy}
\label{subsec:outcomes-close-the-loop-and-govern-autonomy}

Transactions generate time-stamped events that feed an outcomes loop to track latency, safety intercepts, and subgroup disparities. The system earns operational autonomy per task type and dynamically downgrades it via a circuit breaker informed by the task and coordination graph.

\subsection{Vendor copilots as pluggable components}
\label{subsec:evidence-tools-are-pluggable-components-not-competitors}

Organizations can integrate third-party vendor tools as adapters that feed outputs into ClinGuard. The system normalizes all adapter outputs into a common schema, validates them against local governance policies, and compiles them into executable pathway steps with explicit provenance. This allows institutions to swap tools without sacrificing accountability.

\subsection{Four sentinel functions and risk-tier-aware design}
\label{subsec:four-sentinel-functions-and-risk-tier-aware-design}

ClinGuard executes four core functions at runtime (Fig. 2). Monitor watches closure, safety, equity, drift, alert fatigue, subgroup disparities, and incident capture across deployed workflows. Validate ensures every claim binds to source data with timestamps, uncertainty bounds, data-quality flags, and known conflicts. Govern captures agent identity, ownership, and allowed scopes through a deterministic policy engine with risk-tier-stratified thresholds. Escalate routes uncertain or unsafe outputs to appropriate clinical or administrative staff and records the reviewer outcome.

A cross-cutting risk-tier-aware design continuously adjusts these operational thresholds. High-acuity workflows tolerate fewer false negatives, prompting tighter drift thresholds and stricter output validation. Conversely, low-acuity workflows tolerate fewer false positives, prompting ClinGuard to prefer automated revision over block. Therefore, ClinGuard renders tier-conditional decisions and exposes its reasoning within the decision record.

\section{Platform architecture: execution, accountability, outcomes}
\label{sec:platform-architecture-execution-accountability-outcomes}

\subsection{Overview: three supervisory modules plus an external-actor, internal-critic governance loop}
\label{subsec:overview-three-modules-plus-an-actor-critic-governance-loop}

As illustrated in Fig. 1, ClinGuard is organized into three modules providing execution, accountability, and measurement. Module A compiles validated recommendations into workflow-native actions, enacting the govern function via ClinRule predicates. Module B enacts the validate function by binding every claim to source pointers and versioned policies, recording an audit-grade decision record for every transaction, and appending an immutable envelope to a tamper-evident audit ledger. Module C enacts the monitor function to track drift, equity, closure, and alert fatigue. It links actions to closure and downstream outcomes, governing autonomy through real-time telemetry of quality-of-service indicators. Table~\ref{tab:4} specifies the function and location of each module.

\subsubsection{What ``actor-critic governance'' means in ClinGuard}
\label{subsubsec:actor-critic-governance}

Actor-critic governance describes the structural relationship between ClinGuard and the coordinating agent it supervises. The external actor is the language model assistant. It generates proposals, such as drafted orders or referral packets, and is optimized for coverage and throughput. The internal critic is the ClinGuard sentinel module. It evaluates each proposal against fixed, versioned governance artifacts and is optimized for safety and compliance to intercept non-conforming actions.

ClinGuard separates coverage and safety by keeping the critic outside the actor optimization loop. Humans curate the sentinel logic through a manifest registry rather than learning it from artificial intelligence feedback \cite{ref19}. This prevents the critic from drifting to approve unsafe actions and stops the actor from gaming thresholds at deployment time. The system utilizes a closed-loop measurement system rather than a training loop. Updates require regression testing and shadow-mode review before promotion. This measurement loop produces a tier-conditional verdict, an outcomes monitor signal that adjusts system autonomy dynamically, and a decision record exposing both actor and critic reasoning. Section~\ref{subsec:module-c-outcomes-observatory-audit-grade-outcomes-loop} introduces a game-theoretic monitoring signal to audit for abnormal agreement patterns.

\subsubsection{Cross-tier contracts and secure data flow}
\label{subsubsec:cross-plane-contracts}

The three-tier design detailed in Fig. 3 is enforced through typed message contracts. The data tier serves a read-only, purpose-bound block of context information with provenance. The decision tier emits a structured proposal bound to a context hash. Finally, the assurance tier returns a signed authorization over the compiled action hash and the applicable policy identifier versions. This architecture keeps minimum-necessary data flow explicit and reviewable.

\subsection{Coordination transactions: explicit state transitions with closure}
\label{subsec:coordination-transactions-explicit-state-transitions-with-cl}

A coordination transaction advances a pathway-defined task state graph. Each transaction incorporates a trigger event, time-stamped context information, a proposed state transition with required artifacts, the deterministic verdict from ClinGuard, a committed audit artifact, limited-privilege execution, and monitoring to closure. Trace identifiers group related transactions into an episode. Unique transaction identifiers support forensic reconstruction when rework is necessary. Fig. 4 illustrates this transaction lifecycle, and Box 2 summarizes how ClinGuard operationalizes governed transactions.

\begin{tcolorbox}[colback=white,colframe=black,title={Box 2. Sentinel gate sequence for governed coordination transactions}]
The sentinel applies a sequential gate sequence before execution: (1) scope, privacy, and inherited permissions, (2) evidence provenance (no claims without pointers), (3) pathway and policy mapping (prerequisites and prohibitions), (4) completeness and internal consistency for execution, and (5) risk tiering and accountability (assigned human owner, sign-off requirements). Gate failures return REVISE directives, or VETO and ESCALATE with reason codes.
\end{tcolorbox}

\subsection{Sentinel critic implementation: gates, predicates, and concept resolvers}
\label{subsec:control-partitioning-llm-services-versus-deterministic-gover}

Figure 5 traces one proposal through the gate sequence of Box~2. The deterministic verdict of the sentinel (APPROVE, REVISE, VETO, ESCALATE) depends strictly on frozen, time-stamped context information and versioned governance artifacts. For ambiguous clinical concepts, such as hemodynamic instability, the sentinel validates that the structured plan is complete, logically consistent, and compliant with explicitly encoded predicates. It prioritizes measurable clinical proxies from vital signs, orders, and medication administrations with institution-defined thresholds, such as a mean arterial pressure below 65 mmHg sustained for more than 15 minutes. When unstructured text requires interpretation, the sentinel applies a conservative concept resolver. Low-confidence inferences are treated as uncertainty and automatically trigger a REVISE or ESCALATE verdict rather than autonomous approval.

Threshold values, such as a confidence cutoff of 0.9, serve as conservative starting defaults rather than universal constants. Every threshold is a versioned, pathway-level parameter tuned per task type, risk tier, and unit context during shadow-mode calibration. Higher thresholds are required for irreversible or high-acuity actions, such as medication updates or discharge against medical advice, while lower thresholds are reserved for low-risk administrative drafts. Calibration jointly tracks alert-fatigue indicators, including override rates and false-escalation rates, against safety indicators like near-miss captures. When an override rate exceeds a pathway-specific ceiling, the threshold is adjusted in shadow mode and re-promoted only after a clinician-adjudicated review. The supplementary appendix (C2) shows the corresponding policy-predicate definition.

Local tuning is constrained by deployment-wide invariants: every state-changing action passes the gate sequence; every claim has a source pointer; unresolved uncertainty cannot yield approval; every verdict is recorded; and every pathway names an owner and escalation route. Numeric cutoffs, latency targets, task risk tiers, and reviewer routing may vary only within declared ranges. The manifest registry rejects specifications that omit an invariant or place a tunable outside its range. Supplementary Appendix~K describes calibration, predicate-acceptance tests, and the full invariant set.

The critic operates via a hybrid pipeline:
\begin{enumerate}
  \item Contract validation ensures required structured fields are present, typed, and non-null.
  \item Rule and ontology validation evaluates policy predicates and pathway clauses over normalized inputs, such as FHIR resources, using compiled, versioned governance manifests keyed by stable policy identifiers.
  \item Optional concept resolvers utilize sandboxed extractors to populate specific features with confidence bounds, routing low-confidence outputs to mandatory escalation.
\end{enumerate}

\FloatBarrier

\subsection{Evidence abstraction: pluggable evidence sources via an evidence-adapter interface}
\label{subsec:evidence-abstraction-pluggable-evidence-sources-via-an-evide}

ClinGuard treats evidence tools as pluggable adapters that normalize heterogeneous outputs into a common evidence-item schema (Supplementary Appendix~A). The platform arbitrates this evidence against local context, including local policy precedence and resource constraints, compiles it into executable pathway steps, and logs it into the decision record to preserve clear provenance and operational accountability.

\subsection{Module A: pathway engine}
\label{subsec:module-a-pathway-execution-engine-clinical-pathway-executor}

Module A provides the runtime realization of the govern pillar (Section~\ref{subsec:four-sentinel-functions-and-risk-tier-aware-design}). It compiles the versioned policy artifacts and pathway specifications into the gated, workflow-native actions that appear within the electronic health record. This module translates clinical guidelines, standard operating procedures, and payer constraints into versioned, exception-aware pathway specifications that define triggers, allowed states, prerequisites, exceptions, risk tiers, and closure criteria (Supplementary Appendix~B). Approved transitions are compiled into workflow-native tasks, such as FHIR tasks, messaging requests, or referral packets, with declared write targets, required sign-offs, and explicit rollback plans (Supplementary Appendix~C).

\subsection{Module B: decision record and audit ledger}
\label{subsec:module-b-tradeoff-ledger-decision-tradeoff-record-and-trustl}

Module B provides the runtime realization of the validate pillar, ensuring every claim made by the coordinator agent binds directly to a source pointer with a timestamp, an uncertainty bound, and a record of conflicting evidence. It materializes decision-making into a signable decision record that details the evidence used, triggered policies, constraints, alternatives considered, the sentinel verdict, and the confirmation chain (Table~\ref{tab:5}). A compact ledger envelope binds this record to cryptographic hashes of the context information, proposal, critique, and action. This structural design provides tamper-evident traceability for incident reconstruction \cite{ref20}. (Concrete field-level schemas for the envelope and the decision record are provided in Supplementary Appendix~H). The context hash also acts as a temporal lock against retroactive gaming; by hashing the transaction input state at time $T_0$ before an action is proposed, the sentinel validates the plan strictly against what the system knew at that specific moment.

\subsubsection{Ledger integrity model and its limits}
\label{subsubsec:ledger-integrity-model}

ClinGuard uses a centralized, append-only ledger within the deploying institution's trust boundary, not a blockchain or distributed ledger \cite{ref44}. Each envelope contains the preceding envelope's cryptographic hash, so changing an earlier entry invalidates the chain that follows \cite{ref43}. The assurance tier also signs each envelope at commit, binding it to the component that wrote it.

The result is tamper evidence, not absolute immutability. Recomputing the chain reveals retroactive edits but cannot alone prevent privileged truncation or replacement of both the record and its verification tooling. ClinGuard therefore requires periodic publication of the chain-head digest to an external witness \cite{ref45}, write-once ledger storage, and a read-only replica available to institutional oversight. Supplementary Appendix~H gives the corresponding fields.

Distributed consensus becomes relevant when parties lack a shared administrative authority, for example across institutions or between a payer and provider. Within one institution, it would add operational cost without strengthening the audit guarantee supplied by the required controls above.

\subsection{Module C: outcomes monitor}
\label{subsec:module-c-outcomes-observatory-audit-grade-outcomes-loop}

Module C realizes the monitor pillar by watching population-level deployment behavior over time, including closure rates, safety intercepts, equity slices, drift, and alert fatigue. The final pillar, escalate, cuts across all modules to route exceptions to a tier-matched human reviewer, such as a clinician, navigator, or pharmacist, and logs the reviewer's ultimate selection.

Module C links actions to closure and downstream outcomes using pathway-specific outcome specifications (Supplementary Appendix~E). The monitor tracks latency distributions, intercept rates, and subgroup disparities in real time \cite{ref21}, governing autonomy through an automated circuit breaker (Fig. 6).

This is a learning health system loop applied to governance: routine operating data change the autonomy the system is permitted, not the model or clinical guideline. Because the change is recorded in versioned governance artifacts, the resulting autonomy decision remains auditable.

To audit for abnormal actor-critic agreement, the system monitors consensus metrics, such as consensus entropy, as a governance alarm. Near-zero rejection regimes trigger immediate review rather than serving as an optimization target. In mature clinical workflows, low consensus entropy is expected once the coordinator agent aligns with local standard operating procedures. ClinGuard therefore treats consensus entropy as a deviation alarm against a unit-specific baseline established during shadow mode. The alarm fires when running entropy diverges materially from this baseline or when the approval rate reaches 1.0 without recent policy version turnover, exposing potential optimization gaming.

\subsubsection{Event processing and circuit-breaker actuation}
\label{subsubsec:event-processing}

The monitor draws on three sources: pathway state transitions raised by the task state graph, delivery and acknowledgement callbacks returned by the limited-privilege connectors, and electronic health record subscription events that ClinGuard did not initiate. These events are normalized to the model in Supplementary Appendix~E and appended to an episode-keyed ordered stream.

Per-transaction signals such as overdue closure, failed delivery, clustered vetoes, or a missing acknowledgement are evaluated on arrival, with a design target of downgrading autonomy within seconds. Population signals, including subgroup disparities, input drift, alert-fatigue indicators, and consensus entropy, are evaluated over rolling windows. Either class can lower autonomy automatically; only the first can stop an action already in flight.

Windowed monitoring uses a cumulative-sum (CUSUM) procedure to detect persistent departures from the unit's shadow-mode baseline while limiting responses to isolated outliers \cite{ref46}. The baseline is re-established when the pathway, policy set, or acting model changes, consistent with continual monitoring guidance for deployed clinical algorithms \cite{ref47}.

\subsection{Silent-failure prevention: revise, veto, rollback, and escalation}
\label{subsec:preventing-silent-failures-revise-veto-rollback-and-escalati}

The sentinel prevents partial-completion silent failures by making failure states explicit. It issues a REVISE directive for missing prerequisites, a VETO and ESCALATE directive for prohibited transitions, and executes retries or rollbacks for reversible failures. It also maintains a kill switch for pathway-level shutdown during incident response. This formulation establishes a self-corrective architecture. The cross-cutting risk-tier-aware design translates into explicit, tier-conditional behavior across all four pillars, controlling alert-fatigue risks through auditable, versioned calibration policies tied to documented institutional review.

\subsection{Runtime environment, packaging, and deployment}
\label{subsec:runtime-environment-packaging-and-deployment}

A portable supervisory layer requires explicit runtime as well as policy controls. The following are architectural requirements; Table~\ref{tab:evidence} identifies the reference implementation as a proof of concept rather than a production deployment.

\paragraph{Where the system runs.} ClinGuard is designed to run within the deploying institution's trust boundary, in the same administrative and network domain as the electronic health record. The data tier serves purpose-bound context; the assurance tier holds signing keys and the ledger. This placement keeps patient data and the oversight replica under institutional control.

\paragraph{How the pathway engine executes work.} The engine interprets versioned declarative pathway specifications (Supplementary Appendix~B), each with a stable identifier, owner, semantic version, and testable acceptance criteria. It is not a general-purpose scripting or low-code host. Evidence adapters and concept resolvers supply bounded extension points behind typed, sandboxed contracts; their outputs remain unverified until the sentinel checks them.

\paragraph{Packaging and dependency management.} Components are packaged as container images with pinned dependencies and recorded build provenance \cite{ref48}. Each transaction records the evaluating sentinel's image digest with the applicable policy and pathway versions, allowing an auditor to identify the exact evaluator that produced the verdict.

\paragraph{Configuration and promotion.} Governance artifacts and environment settings are version-controlled text promoted through the manifest registry. Policy, threshold, and pathway changes require review and regression testing against stored transactions in both directions: newly allowed and newly blocked. Supplementary Appendix~K gives the acceptance tests and promotion rule.

\paragraph{What keeps the layer model-agnostic.} A single adapter gives the coordinating agent purpose-bound context and accepts a structured proposal; the agent has no execution privilege. Replacing the model changes the adapter implementation, not the sentinel's governance artifacts, ledger, or autonomy ladder. Model substitution still requires regression testing and a new recorded component version.

\section{Operationalization in clinics: integration, governance, and autonomy}
\label{sec:operationalization-in-clinics-integration-governance-and-aut}

\subsection{Integration and knowledge grounding}
\label{subsec:integration-and-knowledge-grounding}

ClinGuard integrates directly with electronic health record infrastructure via SMART-on-FHIR \cite{ref18}. Read operations remain purpose-bound and logged into context information, while write-backs occur only after assurance-tier authorization over an action hash. The coordinator agent treats each clinical claim as a pointer-backed statement. The sentinel rejects unsupported claims and locks reviews to a time-stamped snapshot to eliminate moving-target ambiguity during clinical validation.

\subsection{Governance artifacts and ownership}
\label{subsec:governance-artifacts-and-ownership}

Governance is encoded as versioned artifacts, including atomic policies for gate checks and escalation rules, pathway specifications detailing state machines, closure criteria, and latency targets, and monitoring definitions. Each rule carries a stable policy identifier, semantic version, accountable owner, approvers, and machine-testable acceptance criteria. Artifacts are managed in a manifest registry with explicit change logs and rollback handles to enable reproducible rollouts across clinical units and vendors. Updates undergo regression testing and promotion through shadow mode before enabling higher autonomy, ensuring that ambiguous or conflicting clauses are surfaced as explicit review items rather than being silently encoded.

\subsection{Alignment with regulatory guidance for clinical artificial intelligence}
\label{subsec:alignment-with-ruaih}

Recent guidance from health-system, regulatory, and standards bodies converges on matching requirements, specifically governance structures, privacy and security controls, ongoing monitoring, and risk and bias management \cite{ref31,ref32}. ClinGuard operationalizes these elements at transaction time via policy-gated authorization, in the ledger through reconstructable traceability, and post-deployment using outcomes and equity monitoring with downgrade triggers. The system does not replace organizational responsibility; instead, it provides the exact instrumentation needed to enforce it.

Beyond the Joint Commission and Coalition for Health AI guidance \cite{ref32}, ClinGuard intentionally aligns with the broader regulatory landscape driving procurement decisions for clinical artificial intelligence. The U.S.\ Food and Drug Administration's Predetermined Change Control Plan for artificial-intelligence-enabled device software \cite{ref34} requires sponsors to enumerate authorized model and data changes before shipping, binding execution to a specific declared model and policy version. The ClinRule policy engine materializes this discipline at runtime. Every executable rule carries a stable policy identifier and semantic version, every pathway carries a frozen specification version, and every decision record binds by hash to the exact policy versions in force at the moment of decision.

The FDA's 2025 guidance on Artificial Intelligence-Enabled Device Software Functions \cite{ref35} establishes expectations for AI-enabled medical-device software through the total product lifecycle, requiring explicit lifecycle management of model and data changes, ongoing performance monitoring, and bias evaluation across subgroups. The typed cross-tier contracts, append-only ledger, and outcome-linked monitoring within ClinGuard map directly onto these requirements. Furthermore, the NIST AI Risk Management Framework \cite{ref36} structures organizational responsibility around four core functions: Govern, Map, Measure, and Manage, each of which ClinGuard's three internal modules (A-C) realize. The NIST SP 800-53 security controls \cite{ref37} for access control, audit, and system integrity correspond directly to the limited-privilege connectors and the audit-ledger envelope. Finally, the World Health Organization guidance on the ethics and regulation of artificial intelligence for health \cite{ref38,ref39} provides the international counterpart to these standards. The European Union's Artificial Intelligence Act (Regulation (EU) 2024/1689) \cite{ref42} sets parallel obligations for high-risk AI systems, human oversight, automatic logging and traceability, transparency, and risk management, which ClinGuard meets through tier-matched escalation and sign-offs, the append-only audit ledger, plain-text decision records, and the outcomes monitor, respectively. Table~\ref{tab:6} summarizes this entire regulatory mapping, and Supplementary Appendix~G expands on the seven-element checklist for the Responsible Use of AI in Healthcare. These regimes also leave technical gaps that ClinGuard is built to close: it audits at transaction time rather than after the fact, provides an external kill switch through the circuit breaker, and, through the federated registry of governance manifests described in Section~\ref{sec:outlook-from-pilot-to-platform}, moves toward the shared standards and infrastructure that no single regulation yet supplies.

\subsection{Supervised dynamic autonomy and circuit breaker}
\label{subsec:supervised-dynamic-autonomy-and-circuit-breaker}

The framework implements a tiered autonomy model adapted from SAE automotive driving-automation levels \cite{ref22}. Autonomy is granted dynamically per task type, pathway version, and specific unit context. As illustrated in Fig. 6, Level 0 represents a shadow mode that drafts but does not execute actions, Level 1 functions as a copilot requiring human approval for execution, and Level 2 enables bounded autonomy reserved strictly for defined, low-risk administrative actions. If system telemetry degrades due to near-misses, latency spikes, fairness breaches, or abnormal agreement patterns, the sentinel triggers a circuit breaker that automatically downgrades the autonomy level and reroutes the remaining workflow tasks to human review.

\section{Pilot domain and broader application family}
\label{sec:pilot-domain-high-reliability-discharge-planning-and-snf-pla}

\subsection{Why the family of coordination workflows is the right wedge}
\label{subsec:why-discharge-planning-is-a-high-leverage-wedge}

Coordination workflows that depend on routing, acknowledgment, and closure share a common vulnerability: accumulated omissions across handoffs, systemic drift, and silent state loss. While skilled nursing facility placement represents a stringent, high-volume, and well-measured baseline \cite{ref23,ref24,ref25}, these same challenges occur across the broader workflow family. These include hospital-to-home transitions where missing follow-up drives readmissions; specialty referral closures where patients are lost between primary care and specialists \cite{ref28,ref29}; prior authorizations and medication refills where administrative burdens delay care \cite{ref8,ref30}; transitional heart failure care where clinicians omit guideline-directed therapies; and rural health handoffs where vendor and bandwidth heterogeneity amplify silent failures. Supplementary Appendix~J catalogs the specific evaluation metrics, critical-path sentinel mechanisms, and key performance indicators for each of these six scenarios as a planned evaluation roadmap.

\subsection{Worked example: skilled-nursing-facility placement}
\label{subsec:use-case-smart-discharge-for-snf-placement}

We use skilled nursing facility placement as a worked example because it is thoroughly documented and operationally measurable. Upon a discharge trigger, the coordinator agent assembles a discharge bundle and drafts facility-specific referral packets. ClinGuard then validates completeness, payer constraints, and capability matches before authorizing outreach through limited-privilege connectors. The system tracks the episode as a state machine (sent, received, reviewing, and accepted or declined) to enable deadline-aware escalation and deterministic closure. Fig. 7 contrasts this worked example with the five other coordination workflows across key operational dimensions: acuity, latency targets, and the primary sentinel pillar on the critical path. Supplementary Figure~S5 provides a detailed timeline for an individual transaction.

\subsection{Planned stepped-wedge trial (Duke Health)}
\label{subsec:planned-stepped-wedge-trial-duke-health}

Because no result reported here comes from live clinical operation, the framework's clinical claims remain to be tested prospectively. We propose a pragmatic, stepped-wedge cluster-randomized trial across inpatient discharge units for that purpose \cite{ref27}. Clinical units will begin with usual care and cross over at randomized intervals to platform-assisted coordination. Each cluster initially runs in Level 0 shadow mode to calibrate gates and establish baseline latencies, then transitions to Level 1 copilot mode, which requires human sign-off for all externally facing actions. The primary outcomes are event-timestamped latency to placement closure and transaction rework rates. Secondary outcomes include safety intercepts, near-misses, and equity-stratified performance across patient language, insurance type, clinical site, and specific unit. A multidisciplinary safety and operations board will oversee pre-specified downgrade and kill-switch triggers, working alongside the Algorithm-Based Clinical Decision Support (ABCDS) oversight framework at Duke University.

\section{Bench-scale characterization of the governance plane}
\label{sec:bench-evaluation}

We tested two engineering properties of the reference implementation: incremental latency under load and response to injected transaction faults. This predeployment characterization does not evaluate clinical benefit.

\paragraph{Evaluation design.} Load was generated with Locust 2.34, which drove the production-facing transaction endpoint over HTTP with no test-only code path. The MIMIC-IV-FHIR concurrency arm ran 1, 2, 4, 8, 16, and 32 clients for 300 seconds per level (Table~\ref{tab:bench}). A second arm compressed the hourly distribution of 275 encounter starts into a 720-second surge, weighted 1.8$\times$ at the 07:00, 15:00, and 23:00 shift boundaries, peaking at 24 clients. Stage timers separately recorded context assembly, evidence retrieval, model proposal, sentinel critique, autonomy resolution, ledger append, and escalation drafting. For fault testing, we froze one proposal--evidence--context triple per patient, injected five mutations, and replayed each mutant against its paired control: 200 Synthea-derived proposals and 100 MIMIC-IV-FHIR proposals. The coordinating agent was \texttt{gpt-5.4-mini-2026-03-17} throughout.

\paragraph{Load and latency.} Throughput increased from 2.0 to 66.4 transactions per minute, with no failure or saturation through 32 clients (Table~\ref{tab:bench}). End-to-end median latency was 23.6--27.6 seconds. The measured governance stages contributed medians of 78.5--111.4 milliseconds and 0.74--0.98\% of transaction time (Fig. 8a). At 32 clients, median times were 26.8 seconds for the model call, 51.7 milliseconds for context assembly, 13.0 for sentinel critique, and 9.1 for ledger append. The surge arm completed 198 transactions without failure; median governance latency was 92.9 milliseconds.

\paragraph{Interception of non-conforming transactions.} Five classes of mutation were applied: removing a prerequisite, fabricating a policy identifier, crossing an unmet checkpoint, using a stale context hash, and reporting partial execution as complete. Each class represents a failure mode documented in care coordination: transfers proceeding without required documentation \cite{ref10,ref54}; fabricated guideline or policy identifiers, a known failure of language models \cite{ref12,ref13}; workflow steps performed before their preconditions have been satisfied; actions based on patient information that is no longer current \cite{ref9,ref55}; and referral or follow-up loops marked as closed before the underlying work is complete \cite{ref28,ref29}. Interception required the expected check to fire on the mutant but not its paired control. Pooled interception was 91.1\% in Synthea (1000 mutants; Wilson 95\% confidence interval, 89.2--92.7\%) and 87.4\% in MIMIC-IV-FHIR (500 mutants; 84.2--90.0\%) \cite{ref52,ref53}. Four classes were intercepted in 99--100\% of Synthea cases (Fig. 8b). Missing prerequisites yielded 56.5\% paired interception because the required preadmission documentation was absent from every control; nevertheless, the injected fault changed the verdict from revision to hard block in 95\% of cases. These results verify the injected mechanisms, not sensitivity to naturally occurring errors. These classes were derived from the gate contract rather than from observed incidents, so an independently derived fault set is still needed. The shadow phase of the planned trial will capture naturally occurring faults for clinician labeling, while institutional incident reports may provide an additional source of failure cases. Supplementary Appendix~L4 maps each class to the corresponding failure mode documented in the literature.

\paragraph{Audit and actuation cost.} Median ledger append remained 0.032--0.039 milliseconds from 100 to 100,000 entries. Verifying a 100,000-entry chain took 656 milliseconds with no integrity error. Kill-switch and veto-rate-breaker actuation took 0.131 and 0.472 milliseconds, respectively.

\paragraph{Scope of inference.} Absolute latency reflects one workstation and a hosted model interface. One model backbone was used, so model substitution was not tested. No transaction reached approval; contributing causes included missing Preadmission Screening and Resident Review (PASRR) forms, coverage records, and point-of-contact records. Supplementary Appendix~L reports the supporting results, control-arm check specificity, and model usage.

\section{Outlook: from pilot to platform}
\label{sec:outlook-from-pilot-to-platform}

A successful pilot represents an initial milestone rather than the final objective. As ClinGuard expands from a single institution and workflow to multi-site, multi-domain coordination, three platform-level shifts emerge.

\paragraph{Governance artifacts as durable infrastructure.} The audit ledger, decision-record schema, policy registry, and outcome specifications outlive individual model backends. If an institution replaces a coordinating agent, the same versioned pathway specifications and ClinRule predicates supervise the new model without requiring infrastructure re-engineering. Over a multi-year horizon, these governance artifacts become the durable institutional asset, independent of specific model releases.

\paragraph{Federation as a compliance commons.} Pathway specifications and policy manifests authored at one institution can be published as standardized, versioned community artifacts. A federated registry of governance manifests allows smaller institutions to adopt existing policies rather than writing them from scratch. This turns an isolated compliance burden into a shared resource, enabling clinical AI policies from one health system to be transparently inspected, contested, and reused by another.

\paragraph{Procurement based on policy inspection.} With a fixed supervisory substrate, healthcare systems can evaluate vendors by interrogating the exact policy manifests an agent must satisfy, the audit traces it produces, and its performance on the autonomy ladder during shadow mode, rather than relying on unverified performance claims. ClinGuard is designed to integrate with external assurance environments, including institutional algorithm oversight programs, the emerging Health AI Assurance Laboratories network \cite{ref40}, and regulatory post-market surveillance frameworks. Under this model, third-party verification of agent behavior under documented governance becomes a prerequisite for procurement and ongoing operation.

\section{Limitations}
\label{sec:limitations-and-open-challenges}

A governed architecture mitigates but does not eliminate deployment risk. Data quality constraints, rigid policies, and human factors such as automation bias and alert fatigue can still degrade performance. Furthermore, certain predicates rely on features derived from natural language text; these inferences remain probabilistic and require explicit uncertainty tracking and escalation. Privileged execution surfaces also introduce security and liability concerns, which motivates least-privilege design, continuous monitoring, and rapid rollback capabilities.

A major practical constraint is the current asymmetry in SMART-on-FHIR support across healthcare software ecosystems. While electronic health record vendors broadly standardize read access, write-back permissions such as placing orders, drafting tasks, or transmitting referrals remain highly restricted, vendor-gated, and contractually negotiated. Several vendors expose these capabilities exclusively through proprietary application platforms with independent certification cycles. Although ClinGuard's least-privilege execution model minimizes this operational surface by ensuring only authorized action hashes traverse a connector, the framework cannot overcome vendor-side write-back limitations natively. Therefore, we treat SMART-on-FHIR write-back availability as a deployment-time prerequisite rather than an architectural assumption.

The worked examples in this study utilize the MIMIC-IV Fast Healthcare Interoperability Resources (FHIR) dataset rather than live deployment feeds, meaning transportability to specific commercial systems remains untested. To address this, the planned stepped-wedge trial at Duke Health (Section~\ref{subsec:planned-stepped-wedge-trial-duke-health}) extends evaluation beyond MIMIC-IV-FHIR to live institutional data and workflows.

Bench testing bounded but did not resolve scalability (Section~\ref{sec:bench-evaluation}). Through 32 clients, the deterministic governance stages contributed less than 1\% of transaction time and ledger append did not increase with chain depth to 100,000 entries. These measurements came from one workstation, a hosted model interface, and retrospective corpora. They do not predict latency or reliability under sustained production load, live feeds, or concurrent clinical users.

Heterogeneous electronic health record environments present a separate portability problem. FHIR versions, licensed modules, populated fields, local code systems, and extensions vary by site, while some departmental systems expose no modern interface. Predicates inherit those differences and require site-specific validation even when the policy artifact is portable. Moving the framework to a new institution therefore requires a defined local validation process, not a new installation. First, the institution's FHIR environment is audited to determine which policy prerequisites can be observed reliably. The policy manifest and connector scopes are then adapted and versioned for the local setting. Finally, a frozen retrospective corpus is replayed to establish site-specific baselines for interception, specificity, and latency before the system receives any execution authority. This process follows integration practices reported for deployed clinical machine-learning systems \cite{ref18,ref56} and is detailed in Supplementary Appendix~M.

Versioned governance creates continuing maintenance work. Each pathway, predicate, threshold, and outcome specification needs an owner, review cycle, and retirement decision. Otherwise, encoded policy can drift from practice or superseded artifacts can remain active without accountable ownership. ClinGuard can record that process but cannot staff it; institutions must treat it as an operating cost.

Unexpected conditions become exceptions rather than improvised transitions. A patient leaving against medical advice, a withdrawn facility acceptance, or a duplicate episode is routed to a person. Under-specified pathways may therefore increase escalation volume and alert fatigue \cite{ref51}. ClinGuard also fails closed: if it cannot verify the policy version, reach the assurance tier, or write the ledger, it authorizes no state change. This protects the authorization boundary but makes the supervisor an availability dependency that requires operational contingency planning.

\section*{Acknowledgements}
We acknowledge the job-shadowing opportunity provided by Dr.\ James Nicholas. This project is supported in part by SPARK-ICU R01 grant R01GM139967 (Rishikesan Kamaleswaran).

\section*{Author Contributions}
R.K. and N.J.E. conceived the framework. K.W. formulated the framework, created the associated demos, and wrote the manuscript; A.N. reviewed and polished it. R.M.C. provided expert review on regulatory framing and governance of clinical artificial intelligence. All authors contributed to the article and approved the submission of this manuscript.

\section*{Competing Interests}
K.W., A.N., and R.K. are inventors on pending U.S. provisional patent application No.\ 64/030,784 (``Systems and Methods for Multi-Agentic AI''), filed on 6 April 2026 with Duke University as applicant and assignee, which covers the multi-agentic governance systems and methods described in this manuscript. The other authors declare no competing financial or non-financial interests.

\section*{Figures}

\begin{center}
\includegraphics[width=\textwidth,height=0.6\textheight,keepaspectratio]{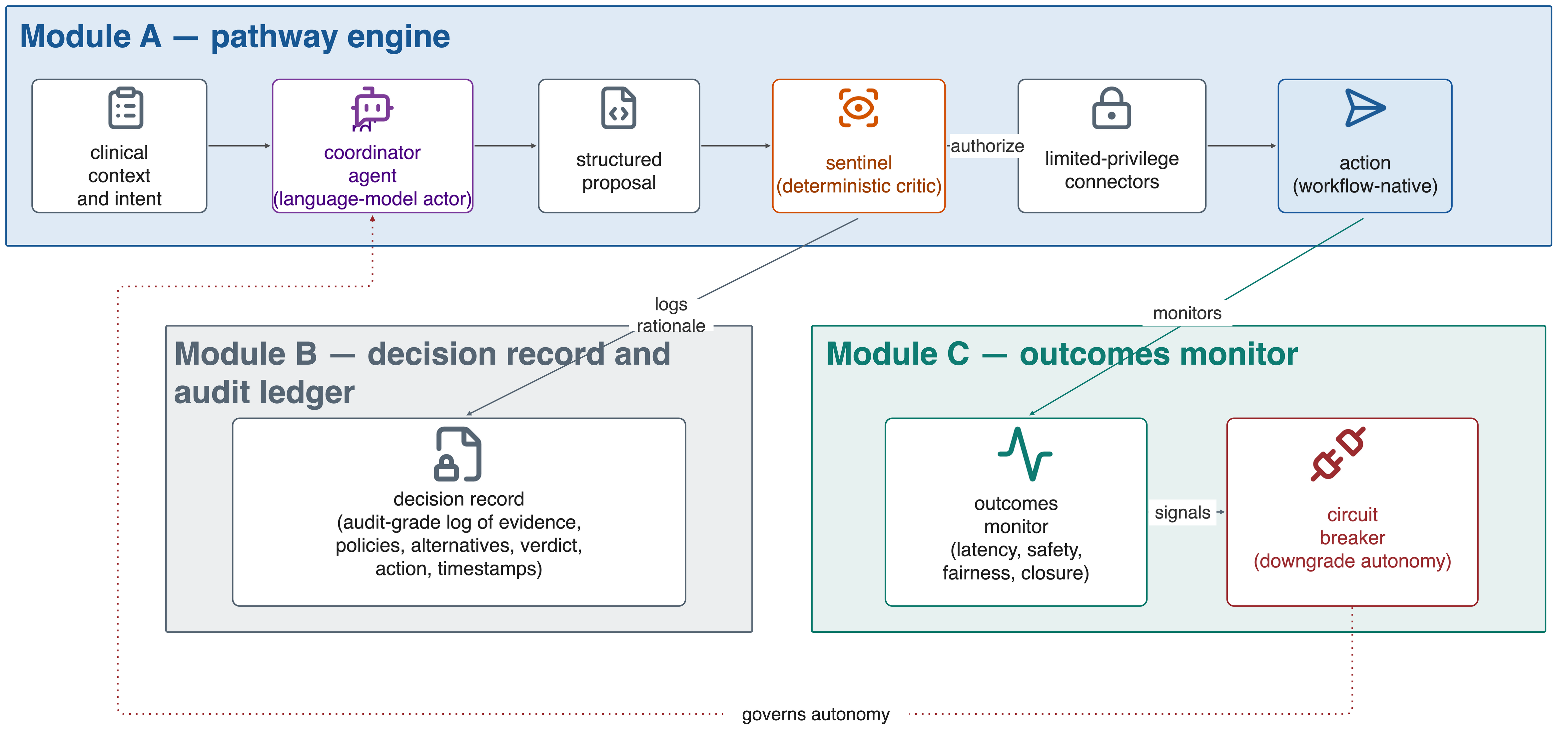}
\end{center}
\smallskip
\noindent\textbf{Figure 1.} \textbf{ClinGuard platform overview: modules and transaction flow.} ClinGuard is an independent supervisory layer organized as three modules: (A) a pathway engine producing workflow-native actions; (B) a ledger that produces an audit-grade decision record for each transaction; and (C) an outcomes monitor that links actions to closure and governs autonomy. The coordinating agent shown above the supervisory layer (a language-model assistant in this rendering) drafts structured proposals and explanations and is rewarded for throughput (task completion and latency reduction). ClinGuard's sentinel validates those proposals against fixed, versioned governance artifacts and is rewarded for safety (interception of non-compliant artifacts and policy adherence), authorizing limited-privilege execution only when gates pass. Module~A enacts the govern pillar, Module~B enacts validate, and Module~C enacts monitor.

\clearpage
\begin{center}
\includegraphics[width=\textwidth,height=0.6\textheight,keepaspectratio]{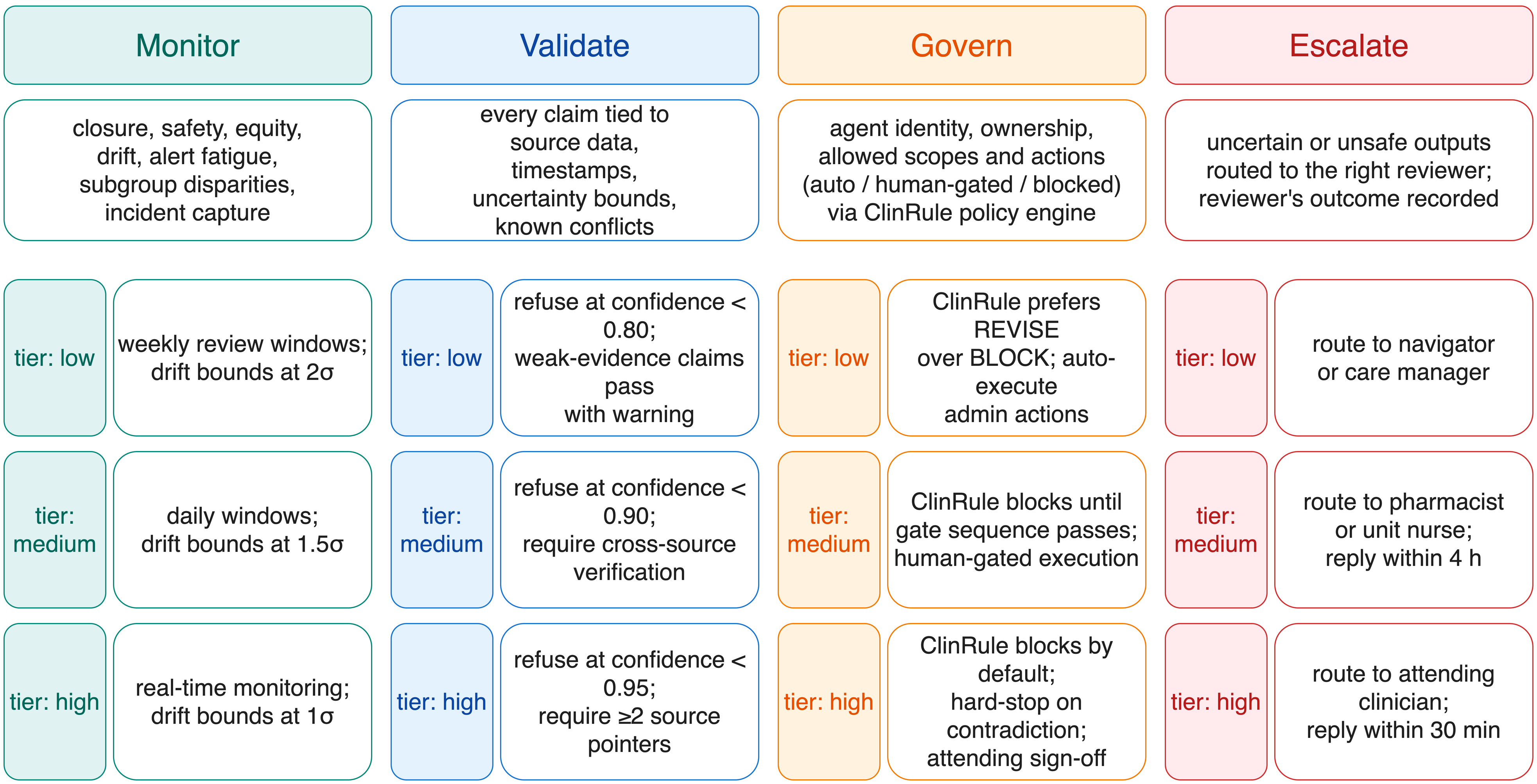}
\end{center}
\smallskip
\noindent\textbf{Figure 2.} \textbf{Four sentinel pillars and risk-tier-aware design.} ClinGuard organizes governance around four sentinel pillars (monitor, validate, govern, escalate) with risk-tier-aware design as a cross-cutting principle (not a fifth pillar). Columns correspond to the four pillars; rows correspond to risk tiers, from low (administrative) through medium (clinical, reversible) to high (acute, irreversible). Each pillar's behavior is conditioned on workflow acuity: higher-acuity tasks tighten monitor's drift bounds and validate's uncertainty thresholds, govern blocks more aggressively, and escalate routes to a higher-credentialed reviewer; lower-acuity tasks invert this. All thresholds are versioned policy artifacts, calibrated per institution and per pathway during shadow-mode deployment.

\clearpage
\begin{center}
\includegraphics[width=\textwidth,height=0.6\textheight,keepaspectratio]{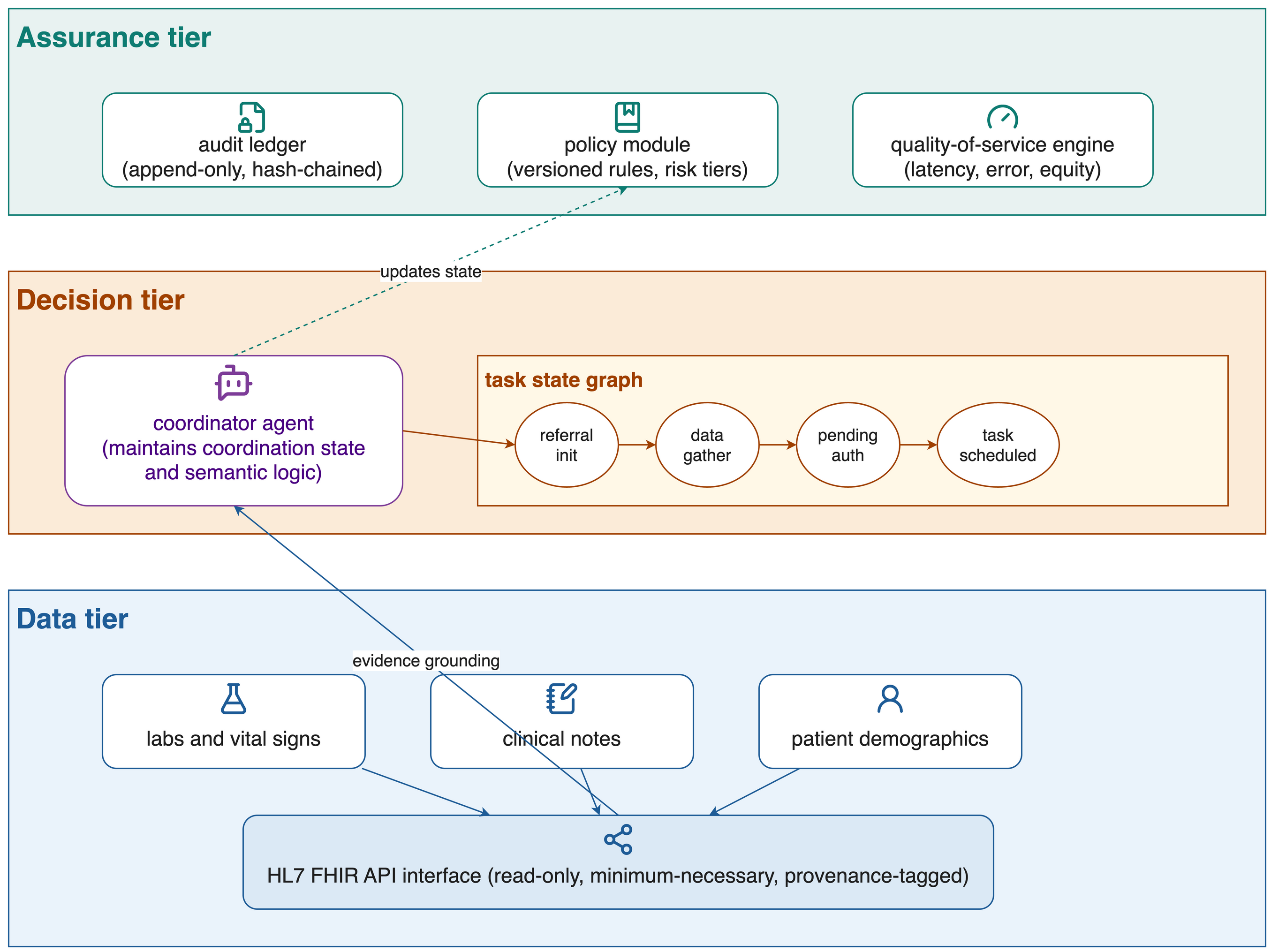}
\end{center}
\smallskip
\noindent\textbf{Figure 3.} \textbf{Three-tier governance architecture: data grounding and tier separation.} Whereas Fig. 1 decomposes ClinGuard by module and transaction flow, this view decomposes the same system by data-access tier. ClinGuard enforces a strict separation of concerns across three operational tiers. Bottom: the data tier handles ingestion and normalization of raw patient data (labs, notes, demographics) via standard HL7 FHIR interfaces, providing the ground truth for evidence grounding. Middle: the decision tier interfaces with the coordinating agent and bounds the agent's outputs to a deterministic task state graph, requiring transitions through discrete stages to ensure process continuity and auditability. Top: the assurance tier functions as the oversight layer, containing the policy module for real-time risk evaluation, the quality-of-service engine for performance tracking, and the audit ledger for immutable logging. Dashed lines indicate control flows, while solid lines indicate data flows.

\clearpage
\begin{center}
\includegraphics[width=\textwidth,height=0.6\textheight,keepaspectratio]{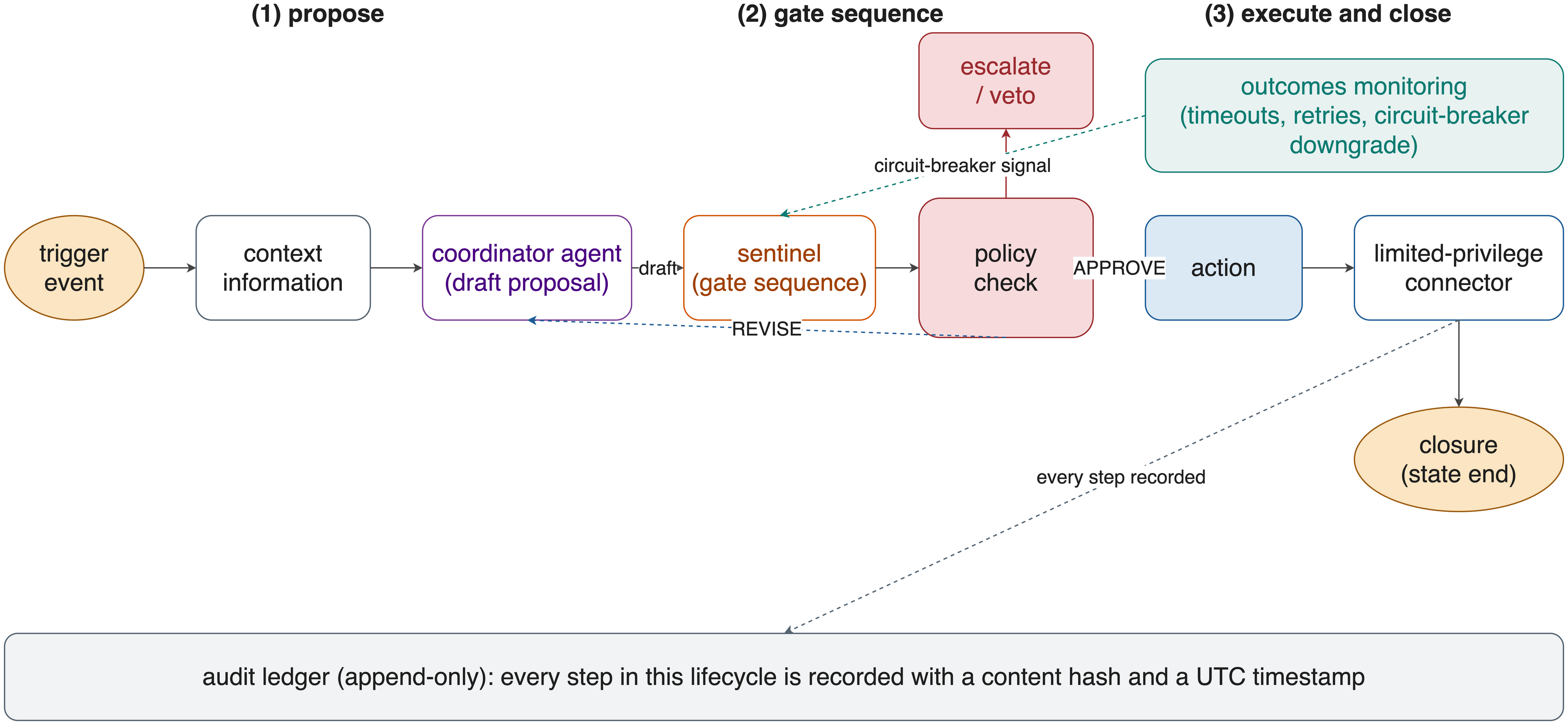}
\end{center}
\smallskip
\noindent\textbf{Figure 4.} \textbf{Coordination transaction lifecycle.} A trigger event initiates retrieval of context information and proposal drafting by the coordinator agent. The sentinel evaluates the proposal through a deterministic gate sequence and issues APPROVE, REVISE, or VETO and ESCALATE. Approved proposals are compiled into actions and executed via limited-privilege connectors. All steps are recorded in an append-only ledger (envelope plus decision record) and monitored to closure with timeouts, retries, and circuit-breaker autonomy control.

\clearpage
\begin{center}
\includegraphics[width=\textwidth,height=0.6\textheight,keepaspectratio]{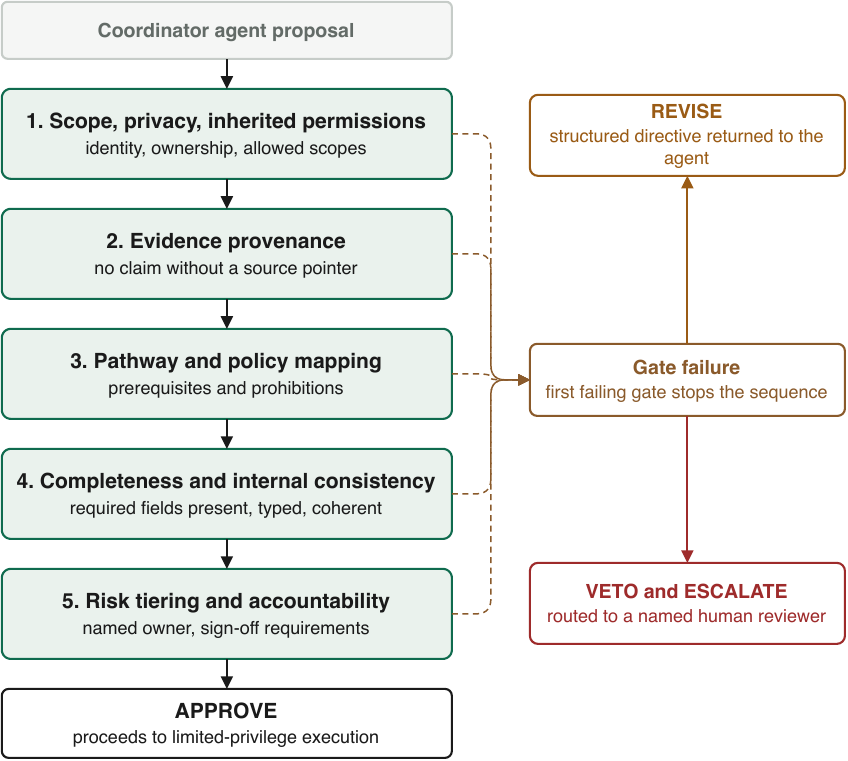}
\end{center}
\smallskip
\noindent\textbf{Figure 5.} \textbf{The sentinel gate sequence applied to a single proposal.} Whereas Fig. 4 shows the transaction lifecycle around the sentinel, this view opens the gate sequence itself. A proposal from the coordinator agent passes through the five gates of Box~2 in order. The sequence stops at the first failing gate, so no gate downstream of a failure is evaluated, and the failure is returned as a structured REVISE directive or, where the violation is disqualifying, as VETO and ESCALATE to a named human reviewer. Only a proposal clearing all five gates reaches execution, and execution is itself limited-privilege. The gates are implemented by the hybrid pipeline described above: contract validation, rule and ontology validation over versioned manifests, and sandboxed concept resolvers with confidence bounds.

\clearpage
\begin{center}
\includegraphics[width=\textwidth,height=0.6\textheight,keepaspectratio]{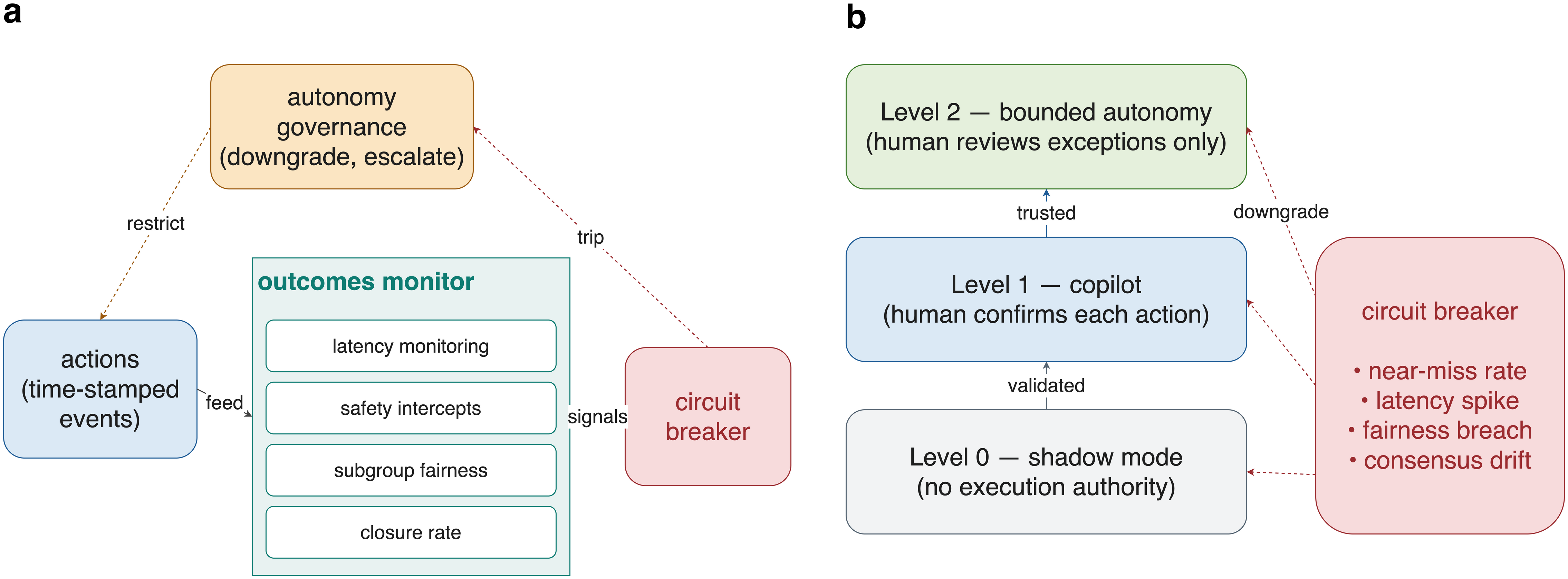}
\end{center}
\smallskip
\noindent\textbf{Figure 6.} \textbf{Outcomes loop and dynamic autonomy.} (a) Actions create time-stamped events that feed the outcomes monitor. Real-time monitoring of latency, safety intercepts, consensus entropy (actor-critic agreement), and subgroup metrics governs autonomy through a circuit breaker and structured escalation, linking deployment behavior to measurable outcomes. (b) Supervised dynamic autonomy ladder. The platform progresses from shadow mode (Level 0) to copilot (Level 1) and bounded autonomy (Level 2) for eligible tasks. Real-time monitoring and ledger-traced performance drive automatic downgrades and escalation when near-misses, latency spikes, consensus drift (deviation from baseline), or fairness breaches occur.

\clearpage
\begin{center}
\includegraphics[width=\textwidth,height=0.6\textheight,keepaspectratio]{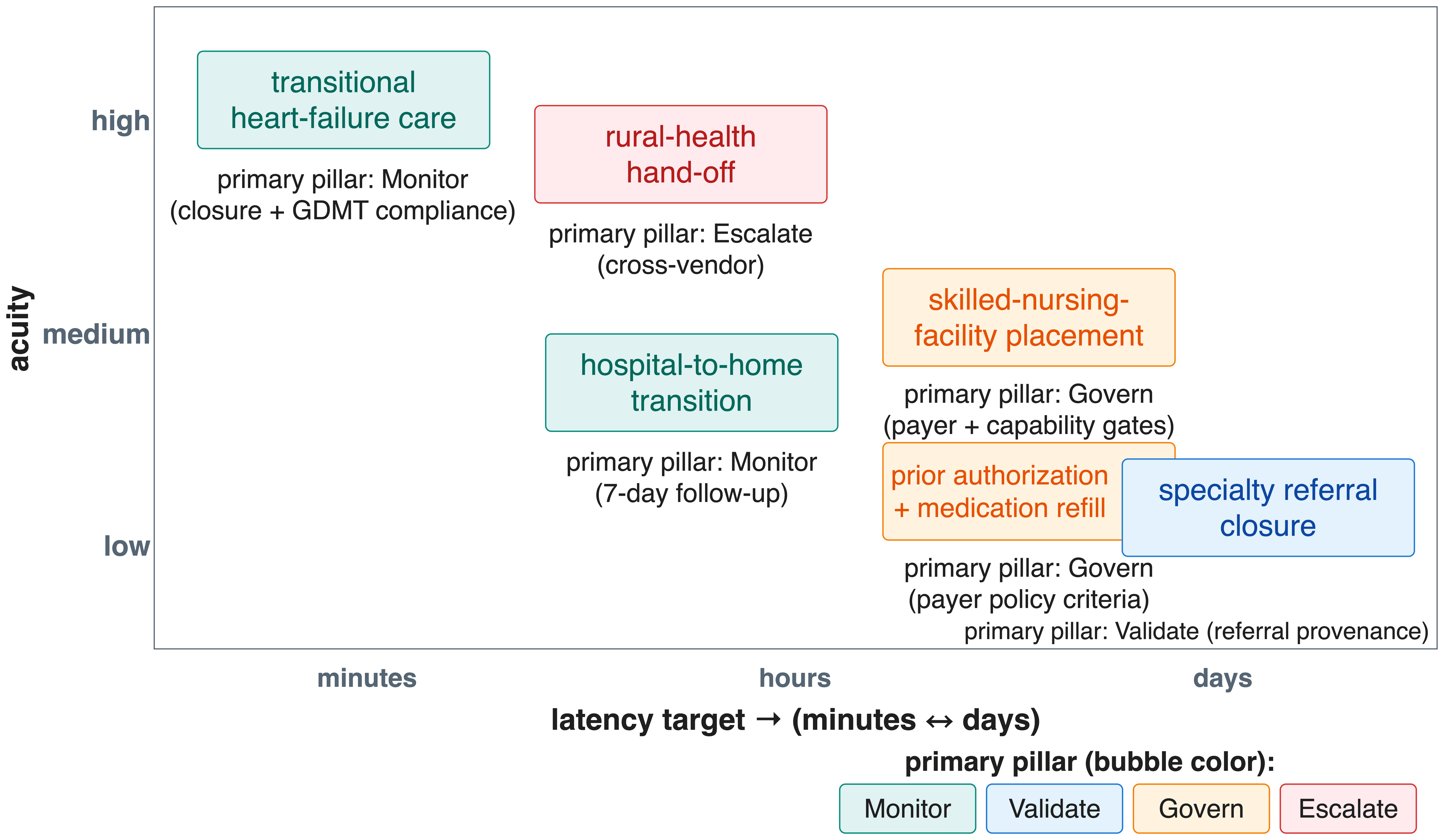}
\end{center}
\smallskip
\noindent\textbf{Figure 7.} \textbf{Six coordination workflows, contrasted on the dimensions that matter.} The same architectural substrate is instantiated across all six coordination workflows. Vertical axis: workflow acuity (low at bottom, high at top). Horizontal axis: typical latency target from trigger to closure (minutes through days). Bubble color marks the primary sentinel pillar on the critical path. The six workflows span the design space rather than repeating a single pattern, motivating the cross-cutting risk-tier-aware design described in Section~\ref{subsec:four-sentinel-functions-and-risk-tier-aware-design}.

\clearpage
\begin{center}
\includegraphics[width=\textwidth,height=0.6\textheight,keepaspectratio]{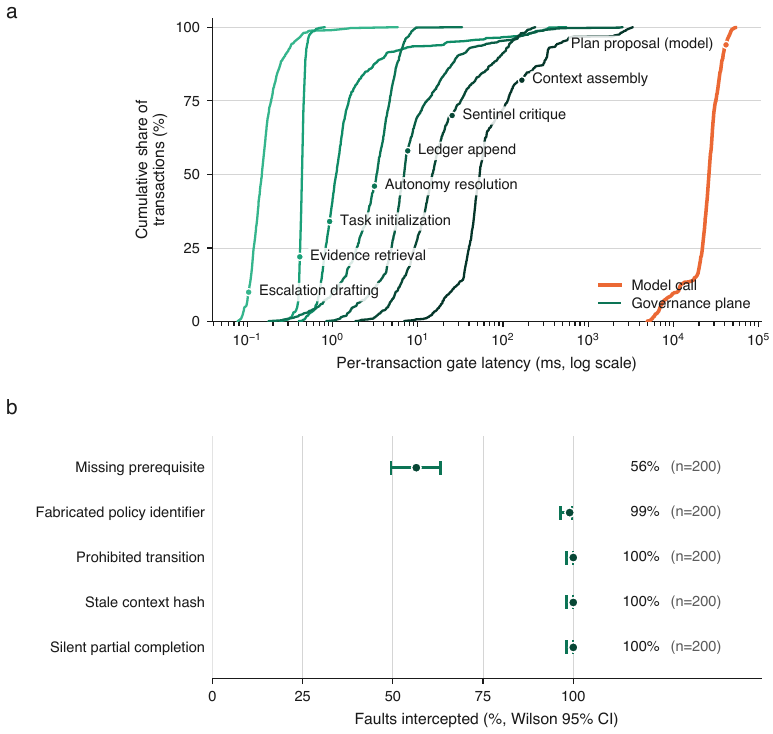}
\end{center}
\smallskip
\noindent\textbf{Figure 8.} Bench-scale characterization of the governance plane. \textbf{a}, Cumulative distribution of per-transaction latency by gate across 674 governed transactions, on a logarithmic axis. The model call sits three orders of magnitude to the right of every deterministic gate; the seven governance gates together resolve well inside the interval occupied by inference alone. \textbf{b}, Interception rate for each class of injected transaction non-conformance, with Wilson 95\% confidence intervals, over 1,000 mutants generated from 200 frozen proposals. The missing-prerequisite interval is lower because the prerequisite policies already fire on every unmutated control in this corpus, not because the check fails to detect the fault; see the text and Supplementary Appendix~L.

\clearpage

\begin{table}[t]
\centering
\caption{Evidence status of ClinGuard capabilities.}
\label{tab:evidence}
\scriptsize
\setlength{\tabcolsep}{4pt}
\begin{tabular}{p{0.30\linewidth} p{0.30\linewidth} p{0.32\linewidth}}
\hline
\textbf{Capability} & \textbf{Evidence status} & \textbf{Where it can be inspected} \\
\hline
Coordination transaction lifecycle and the sentinel gate sequence & Implemented; exercised on retrospective data & Operator console (Appendix~F); worked timeline (Appendix~I) \\
\hline
Decision record and hash-chained audit ledger & Implemented; exercised on retrospective data & Schemas (Appendix~H); worked record (Appendix~D) \\
\hline
Pathway specification and ClinRule policy predicates & Implemented for the skilled-nursing-facility pathway only & Pathway specification (Appendix~B); predicates (Appendix~C2) \\
\hline
Concept resolvers with confidence bounds & Implemented; thresholds are conservative defaults, not yet calibrated on institutional data & Predicate definitions (Appendix~C2); calibration procedure (Appendix~K) \\
\hline
Autonomy ladder and circuit breaker & Implemented as a control surface; downgrade triggers not yet exercised under live operating conditions & Console governance view (Appendix~F); Fig. 6 \\
\hline
Outcomes-monitor metric set (latency, closure, intercepts, equity slices, consensus entropy) & Specified; computed on retrospective examples only & Outcome specification (Appendix~E) \\
\hline
Limited-privilege connector write-back to a commercial electronic health record & Specified, not implemented; contingent on vendor write-back availability & Section~\ref{sec:limitations-and-open-challenges} \\
\hline
The five coordination workflows beyond the worked example & Specified as an evaluation roadmap; not implemented & Appendix~J \\
\hline
Throughput, latency, and behavior under concurrent transaction load & Characterized at bench scale on retrospective data (1--32 concurrent transactions); not tested in production & Section~\ref{sec:bench-evaluation}; Appendix~L \\
\hline
Clinical effectiveness, safety, equity, and workload outcomes & Not evaluated; prospective evaluation planned & Section~\ref{subsec:planned-stepped-wedge-trial-duke-health} \\
\hline
\end{tabular}
\end{table}

\begin{table}[t]
\centering
\caption{Single-agent copilots versus ClinGuard governed coordination transactions.}
\label{tab:1}
\begin{tabular}{p{0.20\linewidth} p{0.38\linewidth} p{0.38\linewidth}}
\hline
\textbf{Dimension} & \textbf{Single-agent copilot (open-loop)} & \textbf{ClinGuard (closed-loop governed transaction)} \\
\hline
Unit of work & A prompt-to-text response or draft artifact; completion is assumed when text is generated. & An explicit state transition that must be validated, executed, and monitored to closure. \\
\hline
State tracking & Implicit and fragmented across inboxes, faxes, portals, and electronic health record notes; dropped tasks are hard to detect. & Explicit task graph and state machine per episode (for example, Ordered $\to$ Packet Sent $\to$ Acknowledged $\to$ Scheduled), persisting across sessions. \\
\hline
Verification & Mostly manual review; the system has no independent mechanism to prevent hallucinations or omissions. & Independent critic validates grounding, policy compliance, safety, and timeliness before execution. \\
\hline
Execution & Either manual copy and paste by staff or automation without a formal safety gate. & Through limited-privilege connectors (FHIR, messaging, portals), authorized only after sentinel approval. \\
\hline
Failure handling & Errors can be plausible and silent (for example, a hallucinated guideline or a missed prerequisite form). & Structured REVISE directives, VETO and ESCALATE pathways, and human handoff for exceptions. \\
\hline
Auditability & Limited traceability; difficult to reconstruct what context the model used. & Audit ledger entries for incident reconstruction: request, context information, proposal, critique, execution, timestamps, and policy identifiers. \\
\hline
Autonomy control & Static: always human-in-the-loop, or ad hoc automation with unclear boundaries that is hard to scale. & Tiered dynamic autonomy (Levels 0--2) earned per task type, with a circuit breaker that downgrades on risk. \\
\hline
Quality and equity monitoring & Rarely systematic; drift and subgroup disparities may go unnoticed. & Quality-of-service engine monitors latency, error rates, near-misses, and fairness slices (for example, insurance, language, race, ethnicity). \\
\hline
\end{tabular}
\end{table}

\begin{table}[t]
\centering
\caption{Positioning of ClinGuard relative to existing governance and supervision approaches.}
\label{tab:2}
\scriptsize
\setlength{\tabcolsep}{3pt}
\renewcommand{\arraystretch}{1.05}
\begin{tabular}{p{0.16\linewidth} p{0.19\linewidth} p{0.19\linewidth} p{0.20\linewidth} p{0.18\linewidth}}
\hline
\textbf{Approach} & \textbf{What it governs} & \textbf{Enforcement point} & \textbf{Unit of audit} & \textbf{Autonomy control} \\
\hline
Human-in-the-loop review & Whether a clinician accepts a suggestion & The person, at the point of use & Whatever the record system happens to log; the rationale is usually absent & Fixed: every action needs a human, whatever its risk \\
\hline
Governance of clinical decision support \cite{ref49} & Which rules exist, who approves them, when they retire & Committee review, before a rule ships & Change logs for rule content, not per-firing rationale & None; it governs rule content, not run-time behavior \\
\hline
Model documentation and registries \cite{ref50} & Disclosure of intended use, training data, and measured performance & Documentation review, at procurement & A static artifact per model version & None \\
\hline
Guardrail libraries for language models & The text a model is permitted to emit & Inside the model-serving path, over text & Per-request logs, typically without clinical provenance & None \\
\hline
Sentinel agents for multi-agent systems \cite{ref33} & Messages exchanged between agents & On the message channel, between agents & Interception logs & None \\
\hline
Monitoring practices from machine-learning operations \cite{ref47} & Population-level performance and drift after deployment & Offline or batched, after execution & Metric time series & Retraining or withdrawal, decided by people on a slow cycle \\
\hline
Evidence platforms & Retrieval and citation of guidance & Advisory, while a draft is being written & A record of what was retrieved & None \\
\hline
Workflow tools (care-management vendors) & Task state visibility and routing & Routing rails, largely outside the record system & Task status history & None \\
\hline
\textbf{ClinGuard} & Each proposed state-changing action & The assurance tier, between proposal and execution & One decision record per transaction, hash-bound to context, proposal, critique, and action & Earned per task type and withdrawn automatically by the circuit breaker \\
\hline
\end{tabular}
\end{table}

\begin{table}[htbp]
\centering
\caption{Role separation between the coordinating agent (the actor) and ClinGuard's sentinel (the critic).}
\label{tab:3}
\scriptsize
\begin{tabular}{p{0.14\linewidth} p{0.41\linewidth} p{0.41\linewidth}}
\hline
\textbf{Dimension} & \textbf{Coordinator agent (actor)} & \textbf{Sentinel (critic)} \\
\hline
Core function &
Interprets clinical intent, assembles context, drafts artifacts, and proposes executable next steps in a task graph. &
Independently validates each proposed step against policy, safety, completeness, and timing; issues APPROVE / REVISE / VETO and escalates exceptions. \\
\hline
Optimization target &
Coverage, usability, and throughput (reduce manual drafting). &
Precision and reliability (prevent unsafe, ungrounded, or incomplete actions from reaching execution). \\
\hline
Grounding and evidence &
Retrieval-augmented generation over electronic health record data and curated guidelines or standard operating procedures; produces structured claims and citations. &
Evidence verification and constraint checking; detects contradictions with electronic health record state or institutional knowledge. \\
\hline
State and closure &
Maintains an explicit state machine for each coordination episode and proposes state transitions toward closure. &
Confirms state transitions, monitors latency, and triggers escalation when a transaction stalls or uncertainty rises. \\
\hline
Interfaces &
FHIR application programming interfaces, scheduling and messaging systems, referral portals, and payer interfaces (via limited-privilege connectors). &
Policy engine, quality-of-service telemetry, privilege management (the circuit breaker), and write access to the audit ledger. \\
\hline
Failure handling &
Regenerates plans and requests missing inputs based on structured critique. &
Produces machine-readable critiques, vetoes unsafe transitions, and enforces human-in-the-loop escalation when needed. \\
\hline
Example task types &
Assemble a skilled-nursing-facility packet; draft a prior authorization; schedule a 7-day follow-up; instantiate a heart-failure checklist and outreach tasks. &
Check pre-admission screening and resident review coverage; verify contraindications and missing documentation; monitor fairness slices and exception rates. \\
\hline
Artifacts logged &
Candidate plan, generated documents and messages, and context-information hashes. &
Verdict, applied policy identifiers, rationale, and performance telemetry (latency, error, equity). \\
\hline
\end{tabular}
\end{table}

\begin{table}[t]
\centering
\caption{Modules of the ClinGuard stack mapped to architectural components and sentinel functions.}
\label{tab:4}
\begin{tabular}{p{0.16\linewidth} p{0.30\linewidth} p{0.30\linewidth} p{0.18\linewidth}}
\hline
\textbf{3-module deliverable} & \textbf{What it does} & \textbf{Where it lives} & \textbf{Sentinel function enacted} \\
\hline
A. Clinical pathway executor & Pathway engine producing \textit{actions} & Mostly decision tier (coordinator agent) plus connectors plus task state graph (Fig. 3) & govern \\
\hline
B. Decision record & Ledger producing a signable \textit{decision record} & Assurance tier (sentinel) plus audit ledger & validate \\
\hline
C. Outcomes loop & Outcomes monitor driving \textit{governed autonomy} & Quality-of-service engine plus telemetry plus circuit breaker (Fig. 6) & monitor \\
\hline
\end{tabular}

\smallskip
{\footnotesize \emph{Note.} The fourth pillar, escalate, is cross-module: any of A, B, or C can invoke it when a tier-relevant exception is detected, routing the transaction to a human reviewer matched to the task's risk tier.}
\end{table}

\begin{table}[t]
\centering
\caption{Audit artifacts: the decision-record payload and its tamper-evident ledger envelope. Concrete field-level schema is in Supplementary Appendix~H.}
\label{tab:5}
\begin{tabular}{p{0.32\linewidth} p{0.62\linewidth}}
\hline
\textbf{Section} & \textbf{What it captures} \\
\hline
Identifiers and trigger & Episode, transaction, pathway specification version, and the trigger event that opened the transaction. \\
\hline
Context information and provenance & A reproducible snapshot of what the system saw at the moment of decision, with pointers and a content hash. \\
\hline
Evidence bundle & Source materials consulted (guidelines, payer policies, local procedures), with strength grade and applicability. \\
\hline
Policies and pathway clauses applied & The rules that fired, their versions, and the predicate results (operationalizing compliance by construction). \\
\hline
Tradeoffs and alternatives & Options considered and rejected, with reason codes, so auditors can reconstruct ``why A and not B''. \\
\hline
Risk tier, required sign-offs, and authorization outcome & Tier-conditional verdict (APPROVE / REVISE / VETO / ESCALATE), revise directives, and the human owners on the confirmation chain. \\
\hline
Action contents, execution status, and monitoring hooks & The bound action sent through the connector, its execution outcome, and links to the outcome specification and equity stratifiers. \\
\hline
Tamper-evident ledger envelope and integrity & A compact envelope binds the record by hash to the context information, proposal, critique, and action contents, with a UTC timestamp and previous-envelope hash for hash-chained, append-only ordering and forensic reconstruction. \\
\hline
Signature and attribution & A signature applied by the assurance tier at commit time, with the identifier and version of the signing key, so an entry is attributable to the component that wrote it and key rotation does not invalidate earlier entries. \\
\hline
External anchoring reference & The identifier and publication time of the most recent chain-head digest written to a witness outside the operating team's control, so truncation of the chain's tail is detectable by an independent reviewer. \\
\hline
\end{tabular}
\end{table}

\begin{table}[t]
\centering
\caption{Regulatory alignment summary: ClinGuard control points by guidance source.}
\label{tab:6}
\begin{tabular}{p{0.26\linewidth} p{0.36\linewidth} p{0.30\linewidth}}
\hline
\textbf{Guidance source} & \textbf{What it requires of clinical artificial intelligence} & \textbf{Concrete ClinGuard control point} \\
\hline
FDA Predetermined Change Control Plan \cite{ref34} & Pre-declare authorized model and data changes; bind execution to a specific declared version. & Policy-identifier versioning, pathway-specification versioning, immutable hashes per decision record. \\
\hline
FDA AI-Enabled Device Software Functions: Lifecycle Management \cite{ref35} & Lifecycle management of model and data changes, performance monitoring, subgroup bias evaluation. & Outcomes monitor, equity stratifiers, audit-ledger replay. \\
\hline
NIST AI Risk Management Framework \cite{ref36} & \textsc{Govern}, \textsc{Map}, \textsc{Measure}, \textsc{Manage} four-function structure. & Module A realizes \textsc{Map} (tasks to pathway specifications); Module B realizes \textsc{Govern} and \textsc{Measure} (via decision record and audit ledger); Module C realizes \textsc{Manage} (via outcomes monitor and circuit breaker). \\
\hline
NIST SP 800-53 \cite{ref37} & Access control, audit, system integrity. & Limited-privilege connectors, audit-ledger envelope, hash-chained provenance. \\
\hline
EU Artificial Intelligence Act (Regulation (EU) 2024/1689) \cite{ref42} & Human oversight, automatic logging and traceability, transparency, and risk management for high-risk AI systems. & Tier-matched escalation and sign-offs, append-only audit-ledger envelope, plain-text decision records, and outcomes monitor. \\
\hline
WHO ethics and regulatory guidance for artificial intelligence in health \cite{ref38,ref39} & Transparency, accountability, equity, governance for clinical artificial intelligence. & Plain-text decision records, named owners per policy identifier, fairness monitoring, transparent escalation. \\
\hline
Joint Commission and Coalition for Health AI Responsible Use \cite{ref32} & Seven operational elements for AI in healthcare (governance, privacy, security, monitoring, safety reporting, bias, education). & Mapped one-to-one in Supplementary Appendix~G. \\
\hline
\end{tabular}
\end{table}

\begin{table}[t]
\centering
\caption{Bench-scale concurrency sweep over MIMIC-IV-FHIR demonstration records. Load is synthetic; the underlying patient records are not. Governance share is the mean fraction of end-to-end transaction time spent outside the model call. Five minutes were run at each level.}
\label{tab:bench}
\small
\setlength{\tabcolsep}{5pt}
\begin{tabular}{rrrrrrrrr}
\hline
\textbf{Conc.} & \textbf{Trans.} & \textbf{Through.} & \multicolumn{2}{c}{\textbf{End-to-end (s)}} & \multicolumn{2}{c}{\textbf{Governance (ms)}} & \textbf{Gov.} & \textbf{Failed} \\
 & & \textbf{(min$^{-1}$)} & \textbf{p50} & \textbf{p95} & \textbf{p50} & \textbf{p95} & \textbf{share} & \\
\hline
1 & 10 & 2.00 & 25.4 & 38.7 & 111.4 & 461.1 & 0.74\% & 0 \\
2 & 23 & 4.60 & 23.6 & 38.8 & 79.0 & 623.2 & 0.90\% & 0 \\
4 & 48 & 9.59 & 23.7 & 39.6 & 78.5 & 798.1 & 0.98\% & 0 \\
8 & 88 & 17.59 & 26.1 & 39.5 & 88.5 & 490.8 & 0.96\% & 0 \\
16 & 173 & 34.57 & 26.1 & 43.0 & 81.5 & 725.6 & 0.97\% & 0 \\
32 & 332 & 66.35 & 27.6 & 44.7 & 85.3 & 676.6 & 0.95\% & 0 \\
\hline
\end{tabular}
\end{table}

\clearpage
\section*{Data Availability}
MIMIC-IV-FHIR is available through PhysioNet under its credentialed-access terms. Synthea-derived records were generated with the publicly available Synthea software. Machine-readable benchmark summaries and transaction-level evaluation outputs underlying Section~\ref{sec:bench-evaluation} are available to editors and reviewers from the corresponding author during peer review and will be deposited in a public repository upon acceptance.

\section*{Code Availability}
The ClinGuard reference implementation, evaluation scripts, and configuration artifacts are available to editors and reviewers from the corresponding author during peer review. Upon acceptance, they will be released in a public repository under a ClinGuard namespace; until then, the development repositories remain private for security and institutional-review reasons.

\section*{Use of Generative AI Tools}
Large language model tools were used during manuscript preparation for brainstorming and language polishing. These tools were not used to generate new scientific analyses, results, or conclusions. All authors reviewed and edited the manuscript and take responsibility for the final content.

\section*{Supplementary information}
The supplementary appendices provide implementation artifacts, worked examples, and supporting results. Appendices A--E give the evidence, pathway, action, decision-record, and outcome contracts. Appendix F shows the operator console; Appendix G maps Responsible Use of AI in Healthcare elements to ClinGuard controls; Appendix H specifies the ledger schemas; Appendix I gives a hash-bound worked timeline; Appendix J lists candidate evaluation metrics; Appendix K documents calibration and invariants; Appendix L reports the supporting bench results; and Appendix M itemizes the validation steps for porting the framework to a new institution.

\clearpage
\setcounter{figure}{0}
\renewcommand{\thefigure}{S\arabic{figure}}
\renewcommand{\theHfigure}{S\arabic{figure}}
\setcounter{table}{0}
\renewcommand{\thetable}{S\arabic{table}}
\renewcommand{\theHtable}{S\arabic{table}}
\begin{center}
{\LARGE\bfseries Supplementary Information}\par\medskip
{\large Supplementary Appendix: Technical Artifacts for ClinGuard}
\end{center}
\bigskip

These appendices accompany the ClinGuard framework paper. ClinGuard is an independent supervisory layer that wraps a coordinating language-model agent, organized around four sentinel functions (\emph{monitor}, \emph{validate}, \emph{govern}, and \emph{escalate}) with risk-tier-aware design as a cross-cutting principle that conditions thresholds and escalation routes on workflow acuity. The schemas, contracts, and worked examples here are grounded in the publicly available MIMIC-IV Fast Healthcare Interoperability Resources dataset and are intended to make the framework testable. Skilled-nursing-facility placement is used as the worked example throughout because it is well-documented and operationally measurable, but the same artifact contracts apply across all six coordination workflows ClinGuard targets.

We use lowercase plain-language phrases (\emph{pathway specification}, \emph{action} and \emph{action bundle}, \emph{decision record}, \emph{ledger}, \emph{outcome specification}, \emph{evidence item}, \emph{context information}) in this appendix's narrative; the equivalent identifiers in code and schemas (\texttt{PathwaySpec}, \texttt{Action Pack}, \texttt{Decision Dossier}, \texttt{TrustLedger}, \texttt{OutcomeSpec}, \texttt{EvidenceItem}, \texttt{ContextSnapshot}) are kept verbatim wherever a downstream implementer would search for them.

\section*{Appendix A. Evidence-adapter interface and evidence-item normalization}
\label{sup:appendix-a-evidence-adapter-interface-and-evidenceitem-norma}

\begin{quote}\textit{Referenced in main manuscript Section on evidence abstraction.}\end{quote}

Evidence tools can be integrated as pluggable inputs via an adapter layer that normalizes outputs into a single \emph{evidence item} type, a normalized, source-agnostic representation of one piece of evidence (a guideline excerpt, a payer policy, a local standard operating procedure) with provenance, strength grade, and applicability. The platform then arbitrates evidence locally, compiles validated recommendations into actions, and logs the basis for action in the decision record. The same adapter pattern applies across all six coordination workflows the main paper enumerates, even when the underlying evidence source differs.

\subsection*{A1. Evidence-adapter interface (illustrative contract)}
\label{supsub:a2-evidence-adapter-interface-illustrative-contract}

Evidence adapters are thin integration layers that (i) accept a query plus minimum-necessary local context, (ii) call a source-specific evidence system (for example, a guideline repository, a payer-policy database, or an external evidence assistant), and (iii) return one or more normalized \texttt{EvidenceItem} objects (Appendix~A2). The interface below is transport-agnostic (in-process module, RPC, or HTTP service); the contract is defined by required operations and request and response schemas.

\begin{codeblock}
# EvidenceAdapter
# -------------------------------------------------------------------
# REQUIRED operations:
#   metadata()         -> AdapterMetadata
#   minimumNecessary() -> MinimumNecessaryPolicy
#   search(req)        -> EvidenceSearchResponse
#   resolve(req)       -> EvidenceItem
#
# HARD requirement: search() and resolve() MUST return normalized EvidenceItem objects.

AdapterMetadata = {
  "adapter_id": "3rdPartyEvidence",          # stable identifier for the adapter/source
  "adapter_version": "2026-01",              # adapter or upstream system version/release
  "evidenceitem_schema_version": "1.0"       # version of the normalized EvidenceItem schema
}

MinimumNecessaryPolicy = {
  "phi_allowed": false,
  "required_context_fields": [
    "jurisdiction",
    "setting",
    "destination"
  ],
  "optional_context_fields": [
    "payer_type",
    "institution_policy_refs",
    "available_services"
  ],
  "prohibited_context_fields": [
    "name", "mrn", "dob", "address", "phone", "email"
  ]
}

EvidenceContext = {
  "jurisdiction": "US-NY",
  "setting": "adult inpatient discharge",
  "destination": "SNF",
  "payer_type": "Medicaid",
  "institution_policy_refs": ["POLICY:discharge.snf.v3"],
  "available_services": ["case_management", "social_work"]
}

EvidenceSearchRequest = {
  "query": "SNF transfer prerequisites and PASRR requirements",
  "context": EvidenceContext,
  "top_k": 5,
  "as_of_utc": "2026-01-18T12:00:00Z"
}

EvidenceResolveRequest = {
  "id": "OE:query:123456",
  "context": EvidenceContext,
  "as_of_utc": "2026-01-18T12:00:00Z"
}

EvidenceSearchResponse = {
  "items": [ EvidenceItem, ... ],
  "raw_trace_id": "opaque-debug-token",
  "retrieved_at_utc": "2026-01-18T12:00:00Z"
}
\end{codeblock}

\subsection*{A2. Example evidence item (normalized object)}
\label{supsub:a1-example-evidenceitem-normalized-object}

\begin{codeblock}
{
  "source": "3rdPartyEvidence",
  "source_version": "2026-01",
  "id": "OE:query:123456",
  "title": "SNF transfer prerequisites and PASRR requirements",
  "strength": {"grade": "Moderate", "method": "source-native"},
  "applicability": {
    "population": "adult inpatient discharge to SNF",
    "exclusions": ["pediatric", "involuntary psychiatric holds"]
  },
  "contraindications": [
    {"type": "policy", "text": "PASRR required for certain diagnoses/conditions"}
  ],
  "key_findings": [
    {"claim": "PASRR must be completed before SNF placement for qualifying patients", "supports": ["link_or_citation"]}
  ],
  "retrieved_at_utc": "2026-01-18T12:00:00Z"
}
\end{codeblock}

\section*{Appendix B. Illustrative pathway specification (YAML)}
\label{sup:appendix-b-illustrative-pathwayspec-yaml}

\begin{quote}\textit{Referenced in main manuscript Section on the pathway engine.}\end{quote}

A pathway specification declares which states a coordination episode can be in, which transitions are allowed, which prerequisites and exceptions apply, and what counts as closure. The example below targets skilled-nursing-facility placement; analogous pathway specifications exist for hospital-to-home, prior authorization, specialty referral, transitional heart-failure, and rural-health hand-off pathways.

\begin{codeblock}
pathway_id: SNF_PLACEMENT
version: 1.3.0
trigger:
  type: EHR_ORDER
  display: Discharge to SNF
states:
  - Ordered
  - PacketDrafted
  - PacketSent
  - Acknowledged
  - Accepted
  - Declined
  - Closed
transitions:
  - from: Ordered
    to: PacketDrafted
    requires:
      - artifact: InsuranceFacesheet
      - artifact: MedicationReconciliation
    policies:
      - policy_id: POL-SNF-004
        version: 2.1
  - from: PacketDrafted
    to: PacketSent
    requires:
      - artifact: PASRR
        when: patient.qualifies_for_pasrr == true
timeouts:
  PacketSent:
    acknowledge_within_hours: 4
    on_timeout: ESCALATE_TO_CASE_MANAGER
risk_tiers:
  PacketSent: LEVEL_2_ELIGIBLE_IF_ALL_GATES_PASS
  ReferralDestinationSelection: LEVEL_1_CONFIRM_REQUIRED
closure:
  - Accepted
  - Declined
  - Closed
\end{codeblock}

\section*{Appendix C. Example action (electronic health record objects)}
\label{sup:appendix-c-example-action-pack-action-bundle-ehr-native-obje}

\begin{quote}\textit{Referenced in main manuscript Sections on the pathway engine and on the sentinel critic implementation.}\end{quote}

\subsection*{C1. Example object}
\label{supsub:c1-example-action-pack-action-bundle}

The action bundle compiles validated recommendations into workflow-native primitives (draft orders, tasks, referrals, messages, and patient education), including explicit guardrails and required human sign-off for higher-risk actions.

\begin{codeblock}
{
  "actionBundleId": "AB-20260117-E456-001",
  "encounterCsn": "E456",
  "patientMrn": "123",
  "trigger": { "source": "cds-hooks", "hook": "order-sign" },
  "actions": [
    {
      "kind": "order_draft",
      "orderType": "consult",
      "name": "Case Management Consult",
      "priority": "stat",
      "reason": "Evaluate post-acute disposition; initiate SNF referral packet",
      "routing": { "target": "OrdersComposer", "mode": "draft" }
    },
    {
      "kind": "order_draft",
      "orderType": "consult",
      "name": "Physical Therapy Eval",
      "priority": "routine",
      "reason": "Deconditioning; assess mobility/ADL needs for disposition",
      "routing": { "target": "OrdersComposer", "mode": "draft" }
    },
    {
      "kind": "order_draft",
      "orderType": "consult",
      "name": "Occupational Therapy Eval",
      "priority": "routine",
      "reason": "Assess ADLs/equipment needs for SNF vs home",
      "routing": { "target": "OrdersComposer", "mode": "draft" }
    },
    {
      "kind": "task",
      "taskType": "nursing",
      "title": "Complete fall risk + mobility screen (baseline) within 8h",
      "assigneePool": "6E Nursing Pool",
      "due": "2026-01-17T22:10:00-05:00",
      "structuredFields": {
        "tool": "MorseFallScale",
        "documentIn": "Flowsheet"
      }
    },
    {
      "kind": "task",
      "taskType": "case_management",
      "title": "Start SNF referral packet prerequisites",
      "assigneePool": "Care Coordination - CM Pool",
      "due": "2026-01-18T14:10:00-05:00",
      "checklist": [
        "Confirm payer + prior auth requirement",
        "Identify preferred facilities (language/geo)",
        "Collect PT/OT notes when available",
        "Confirm isolation needs / special equipment"
      ],
      "structuredFields": {
        "payer": "Medicare Advantage",
        "priorAuthLikely": true,
        "language": "es"
      }
    },
    {
      "kind": "referral_draft",
      "referralType": "post_acute",
      "destination": "SNF",
      "reason": "Functional impairment + limited caregiver support",
      "acceptanceCriteria": [
        "O2 requirement stable <= 3L",
        "No IV vasopressors",
        "PT/OT evaluation completed"
      ],
      "routing": { "target": "ReferralsModule", "mode": "draft" }
    },
    {
      "kind": "appointment_request",
      "clinic": "Primary Care",
      "timeWindowDays": 7,
      "reason": "Post-discharge follow-up after pneumonia admission",
      "routing": { "target": "Scheduling", "mode": "request" }
    },
    {
      "kind": "patient_education",
      "channel": "MyChart",
      "materials": [
        { "id": "EDU-PNA-001", "title": "Pneumonia care and warning signs", "language": "es" },
        { "id": "EDU-FALL-001", "title": "Fall prevention in recovery", "language": "es" }
      ],
      "routing": { "target": "PatientEducation", "mode": "send" }
    },
    {
      "kind": "inbasket_message_draft",
      "toPool": "Care Coordination - CM Pool",
      "subject": "Auto: High-likelihood SNF disposition (AB-20260117-E456-001)",
      "bodyStructured": {
        "whyNow": ["highFallRisk", "limitedCaregiverSupport", "mobilityImpairment"],
        "needed": ["PT/OT evals", "prior auth check", "facility shortlist"],
        "constraints": ["Spanish language preference", "transport barrier"]
      },
      "routing": { "target": "InBasket", "mode": "draft" }
    }
  ],
  "guardrails": {
    "hardStops": [
      { "if": "patientDeclinesPostAcute", "then": "block_referral_draft_and_open_shared_decision_doc" },
      { "if": "hemodynamicInstability", "then": "suppress_SNF_referral_and_escalate_to_attending" }
    ],
    "requiredHumanSignoff": [
      { "role": "Attending", "for": ["referral_draft", "order_draft:Case Management Consult"] },
      { "role": "CaseManager", "for": ["task:Start SNF referral packet prerequisites"] }
    ]
  }
}
\end{codeblock}

\subsection*{C2. Policy-predicate definitions and uncertainty-aware concept resolvers}
\label{supsub:c2-policy-predicate-definitions}

The example action bundle above references guardrail predicates such as \texttt{hemodynamicInstability}. To avoid overclaiming deterministic clinical understanding from unstructured notes, the sentinel treats these predicates as versioned policy artifacts that are deterministic given their inputs and that prefer structured, measurable proxies whenever available. If a predicate cannot be computed from structured sources, a sandboxed concept resolver may propose a value with confidence; low-confidence results trigger REVISE or ESCALATE rather than autonomous execution.

\paragraph{Threshold calibration mirrors main paper Section 4.3.} \textit{[Realizes the \textbf{validate} and \textbf{govern} pillars: validate refuses claims at lower confidence in higher-acuity workflows; govern's ClinRule thresholds are stratified by risk tier.]} Threshold values such as a confidence cutoff of 0.9 are conservative starting defaults, not universal constants. Each threshold is a versioned pathway-level parameter, tuned per task type, risk tier, and unit context during shadow-mode calibration. Higher thresholds (0.95 or above) are expected for irreversible or high-acuity actions; lower thresholds for low-risk administrative drafts. Calibration jointly tracks alert-fatigue indicators (override rate, time to acknowledge, false-escalation rate) against safety indicators (near-miss capture, contradiction interception). When override rate exceeds a pathway-specific ceiling, the threshold is loosened in shadow mode and only re-promoted after a clinician-adjudicated review window. The example below uses 0.9 as that conservative default.

\begin{codeblock}
# Example policy predicate (illustrative): POL-HEMO-001 v1.0.0
# Deterministic given inputs; thresholds are institution-/pathway-defined and versioned.

PolicyPredicate = {
  "policy_id": "POL-HEMO-001",
  "version": "1.0.0",
  "name": "hemodynamicInstability",
  "inputs": [
    "Observation:MAP (time series)",
    "MedicationAdministration:vasopressor_active",
    "Observation:lactate (most recent)"
  ],
  "definition": {
    "unstable_if_any": [
      "MAP < 65 mmHg for > 15 minutes",
      "vasopressor_active == true",
      "lactate >= 4 mmol/L"
    ],
    "unknown_if": [
      "MAP missing AND no vasopressor data available"
    ]
  },
  "on_unknown": "ESCALATE_TO_ATTENDING"
}

# Optional concept resolver (only when structured inputs are missing):
ConceptResolver = {
  "resolver_type": "classifier_or_ontology",
  "examples": ["keyword+SNOMED mapping", "small fine-tuned biomedical encoder"],
  "output": { "label": "unstable|stable|unknown", "confidence": 0.0-1.0 },
  # Conservative default; tuned per pathway and risk tier in shadow-mode calibration:
  "policy": "If confidence < 0.9 then label := unknown"
}
\end{codeblock}

\section*{Appendix D. Decision-record example}
\label{sup:appendix-d-decision-dossier-examples-audit-grade-tradeoff-re}

\begin{quote}\textit{Referenced in main manuscript Section on the ledger and audit-grade outputs.}\end{quote}

The decision record is an audit-grade record of one transaction: the evidence consulted, the policies applied, the alternatives considered, the verdict, the action sent, and the timestamps, readable by clinicians, governance, and incident reviewers.

\subsection*{Worked example: skilled-nursing-facility placement transaction}
\label{supsub:d3-full-example-snf-placement-transaction}

\begin{codeblock}
{
  "trace_id": "7c2f6c7a-34b1-4a0f-8f3b-1a7cf2d0a9cd",
  "transaction_id": "f3a9a2f1-1c2d-4b12-9f90-6d0f06c7d6f0",
  "pathway_id": "SNF_PLACEMENT",
  "pathway_version": "1.3.0",
  "task_type": "SNF_REFERRAL_PACKET",
  "trigger_event": {
    "type": "EHR_ORDER",
    "display": "Discharge to SNF",
    "timestamp_utc": "2026-01-18T14:02:11Z"
  },
  "context_snapshot": {
    "context_hash": "sha256:9c0b...e21a",
    "context_timestamp_utc": "2026-01-18T14:02:45Z",
    "fhir_pointers": [
      "Patient/123",
      "MedicationRequest?patient=123&status=active",
      "Observation?patient=123&code=creatinine&_sort=-date&_count=1",
      "DocumentReference?patient=123&category=discharge"
    ]
  },
  "evidence_items": [
    {
      "source": "LocalSOP",
      "source_version": "2025.11",
      "id": "SOP:CASEMGMT:PASRR",
      "strength": {"grade": "Policy-Mandatory", "method": "institutional"},
      "contraindications": [],
      "key_findings": [
        {"claim": "PASRR must be present before SNF referral packet submission for qualifying patients", "supports": ["SOP:CASEMGMT:PASRR#clause-2"]}
      ],
      "retrieved_at_utc": "2026-01-18T14:02:50Z"
    },
    {
      "source": "3rdPartyEvidence",
      "source_version": "2026-01",
      "id": "OE:query:987654",
      "strength": {"grade": "Moderate", "method": "source-native"},
      "contraindications": [
        {"type": "clinical", "text": "Facility capability mismatch for complex IV infusions"}
      ],
      "key_findings": [
        {"claim": "SNF capability constraints often exclude complex infusions; verify facility capability before transfer", "supports": ["OE:query:987654#item-4"]}
      ],
      "retrieved_at_utc": "2026-01-18T14:03:02Z"
    }
  ],
  "policies_applied": [
    {
      "policy_id": "POL-SNF-004",
      "policy_version": "2.1",
      "predicate_result": "FAIL",
      "reason_codes": ["MISSING_PASRR_DOCUMENT"]
    },
    {
      "policy_id": "POL-MEDREC-001",
      "policy_version": "1.9",
      "predicate_result": "PASS",
      "reason_codes": []
    }
  ],
  "options_considered": [
    {
      "option_id": "Facility_X",
      "pros": ["closest_to_family"],
      "cons": ["capability_unknown"],
      "rejected_reason_codes": ["CAPABILITY_UNVERIFIED"]
    },
    {
      "option_id": "Facility_Y",
      "pros": ["capable_for_iv_abx"],
      "cons": ["out_of_network"],
      "rejected_reason_codes": ["PAYER_OUT_OF_NETWORK"]
    }
  ],
  "conflicts": [
    {
      "type": "policy_vs_plan",
      "description": "Packet submission proposed without PASRR present",
      "resolution": "BLOCK_EXECUTION_REQUIRE_PASRR"
    }
  ],
  "risk_tier": "LEVEL_1_CONFIRM_REQUIRED",
  "required_signoffs": [
    {"role": "CaseManager", "status": "pending", "timestamp_utc": null}
  ],
  "sentinel_verdict": "REVISE",
  "revise_directives": [
    {
      "directive_id": "REV-001",
      "gate": "PathwayPolicyMapping",
      "instruction": "Obtain/attach PASRR form before submitting SNF packet; re-run compile.",
      "missing_artifacts": ["DocumentReference:PASRR"]
    }
  ],
  "action_pack": [
    {
      "action_type": "FHIR_TASK_DRAFT",
      "resource_ref": "Task/draft-001",
      "description": "Create task for PASRR completion and attachment",
      "write_target": "EHR_TASK_POOL"
    }
  ],
  "execution_status": "pending",
  "monitoring_hooks": {
    "outcome_spec_id": "OUT-SNF-PLACEMENT",
    "outcome_spec_version": "0.4",
    "slo_targets": {"time_to_packet_minutes": 5, "time_to_ack_hours": 4},
    "equity_stratifiers": {"preferred_language": "es", "insurance_type": "medicaid"},
    "consensus_rate": 0.82,
    "consensus_entropy": 0.47,
    "drift_alerts": [],
    "bias_audit_window": "monthly",
    "safety_event_refs": [],
    "model_card_ref": "AMC:ClinGuard:2026-01"
  },
  "append_only_integrity": {
    "dossier_hash": "sha256:2e4a...91f3",
    "prev_ledger_hash": "sha256:0000...0000",
    "ledger_timestamp_utc": "2026-01-18T14:03:10Z"
  }
}
\end{codeblock}

\section*{Appendix E. Outcome specification: event-time-stamped metrics}
\label{sup:appendix-e-outcomespec-event-timestamped-metrics-illustrativ}

\begin{quote}\textit{Referenced in main manuscript Section on the outcomes monitor.}\end{quote}

An outcome specification defines which events and metrics to measure for a given pathway: start events, closure events, intercepts, latency targets, and equity stratifiers.

\subsection*{E1. Example event model and metrics}
\label{supsub:e1-example-event-model-and-metrics-for-snf-placement}

Example for skilled-nursing-facility placement:

\begin{enumerate}
\item Start: ``Discharge to SNF order''
\item Required events: PacketSent, FacilityAcknowledged, Acceptance/Decline, AlternateDisposition
\item Closure: Acceptance OR documented alternate disposition
\end{enumerate}
Metrics:

\begin{itemize}
\item Time to closure
\item Rework cycles (missing documents)
\item Contraindication intercept rate
\item Completion gap by language, insurance, site
\item Avoidable length of stay or delay days
\item Consensus entropy and consensus rate (actor--critic agreement; audit trigger when running entropy diverges from baseline)
\item Edge-riding indicators (distribution of latency-target margins; repeated near-timeout completions)
\item Drift monitoring (distribution shift of key inputs; policy-version-change alarms)
\item Safety-event reporting (high-severity intercepts and circuit-breaker events packaged for incident review; de-identified voluntary sharing when appropriate)
\end{itemize}

\noindent\textbf{Event sources and evaluation cadence (main paper, Section 4.7.1).} \textit{[Instruments the \textbf{monitor} pillar.]} The monitor receives pathway state transitions, connector callbacks, and electronic-health-record subscription events. Episode-level metrics are evaluated as events arrive; population metrics use rolling windows. The default window is seven days, matching \texttt{consensus\_entropy\_window} in Appendix~G1, but may be shortened for high-volume units under the controls in Appendix~K1.

\noindent\textbf{Consensus entropy.} Over a rolling window of transactions, let $p_v$ be the empirical frequency of each sentinel verdict category $v\in\{\text{APPROVE, REVISE, VETO, ESCALATE}\}$. Consensus entropy is the Shannon entropy $H=-\sum_v p_v\log p_v$ (low when one outcome dominates, including ``always approve'').

\noindent\textbf{Healthcare baseline interpretation (mirrors main paper Section 4.7).} \textit{[Instruments the \textbf{monitor} pillar: drift, alert fatigue, and consensus-pattern surveillance, with deviation thresholds tightened in higher-acuity workflows.]} In a mature clinical workflow, low consensus entropy is not by itself a failure signal. Once the coordinator agent is well aligned to local standard operating procedures, near-uniform sentinel approval is the expected and desired regime; it reflects fit between the agent and the institution, not collapse. ClinGuard therefore treats consensus entropy as a deviation alarm against a unit-specific, pathway-versioned baseline rather than as an absolute threshold. The baseline is established during shadow mode and recalibrated whenever the pathway specification, policy-identifier set, or acting model changes. The alarm fires when running entropy diverges materially from the established baseline (in either direction), or when approval rate hits 1.0 without recent policy-version turnover, the latter being the gaming pattern the metric is designed to catch.

\subsection*{E2. Outcome-driven autonomy governance}
\label{supsub:e2-outcome-driven-autonomy-governance-summary}

\begin{enumerate}
\item Grant autonomy per (task type $\times$ pathway version $\times$ unit context), not per task type alone.
\item Require stable safety-intercept rates, low rollback frequency, and adherence to latency service-level objectives.
\item Scale monitoring frequency and escalation thresholds to proximity to patient-care decisions (risk-based surveillance rather than a single global monitoring cadence).
\item Monitor subgroup gaps, consensus entropy (deviation alarm), and latency targets, and downgrade autonomy via a circuit breaker when thresholds are breached.
\end{enumerate}

\clearpage

\section*{Appendix F. Operator console (tasks, governance, auditability) of the demo}
\label{sup:appendix-h-demo-operator-console}

\begin{quote}\textit{Supplement to main manuscript.}\end{quote}

\begin{figure}[p]
\centering
\includegraphics[width=0.98\linewidth]{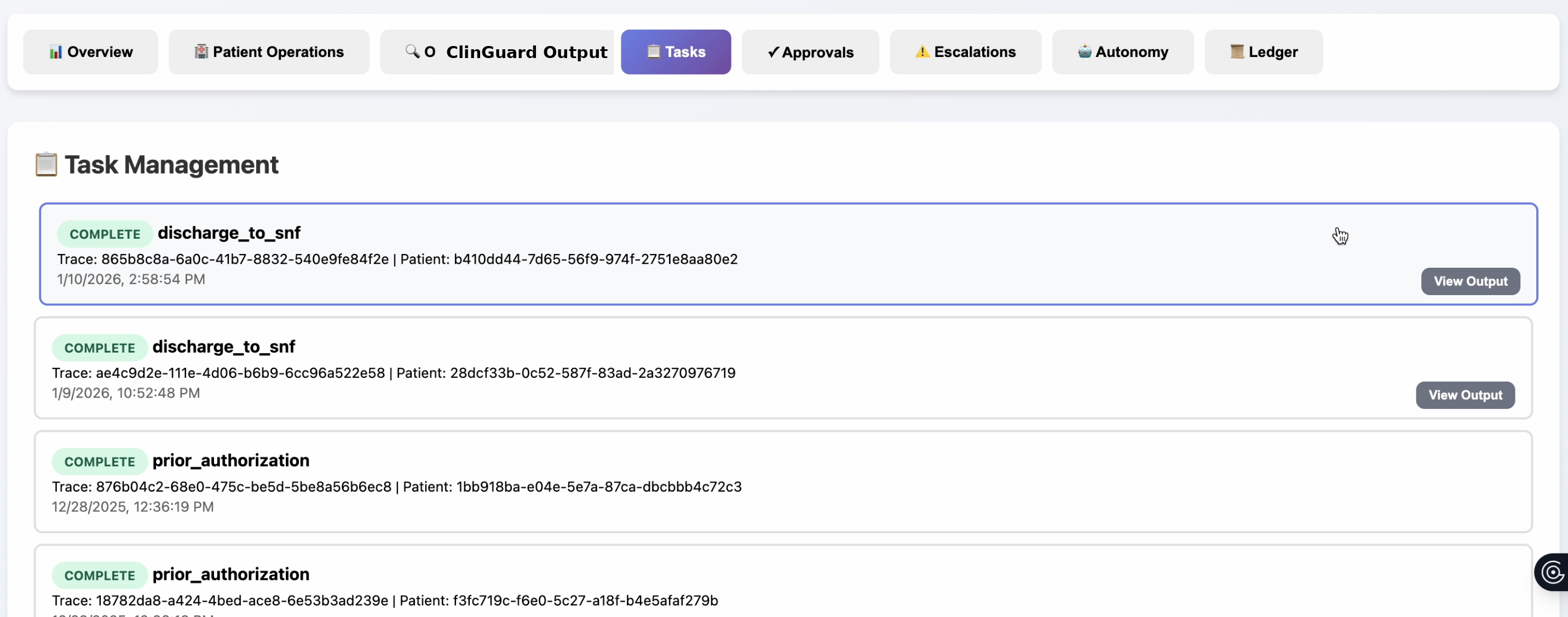}
\caption{\textbf{Operator console, task-management view.} Coordinator-proposed actions operationalized as trackable, human-reviewable work items, with episodes requiring escalation flagged.}
\label{fig:sup-console-lists}
\end{figure}

\begin{figure}[p]
\centering
\includegraphics[width=0.98\linewidth]{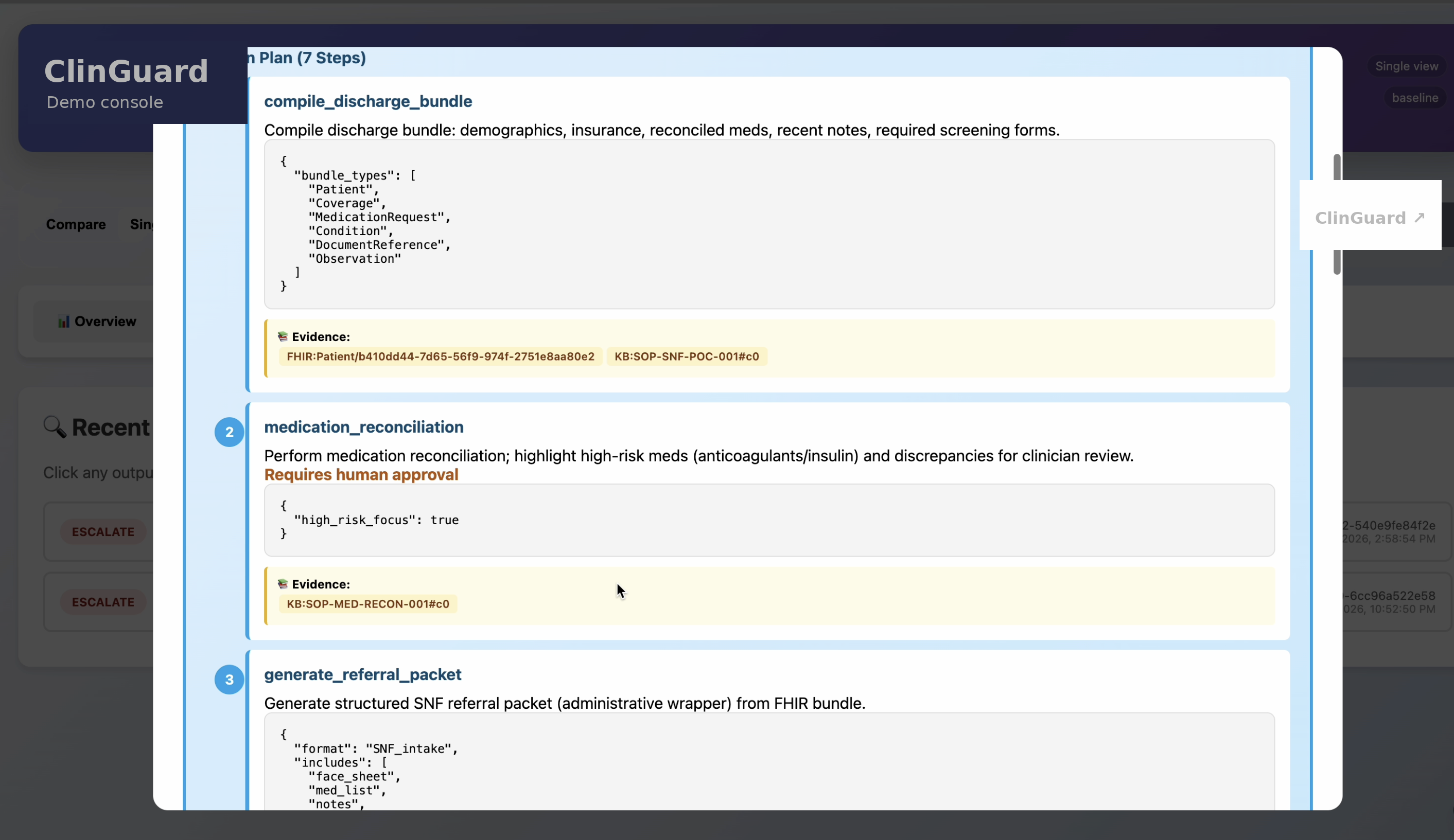}
\caption{\textbf{Structured action plan for a skilled-nursing-facility placement transaction.} Action steps with attached evidence pointers and explicit human-approval gates for high-risk actions.}
\label{fig:sup-console-action-plan}
\end{figure}

\begin{figure}[p]
\centering
\includegraphics[width=0.98\linewidth]{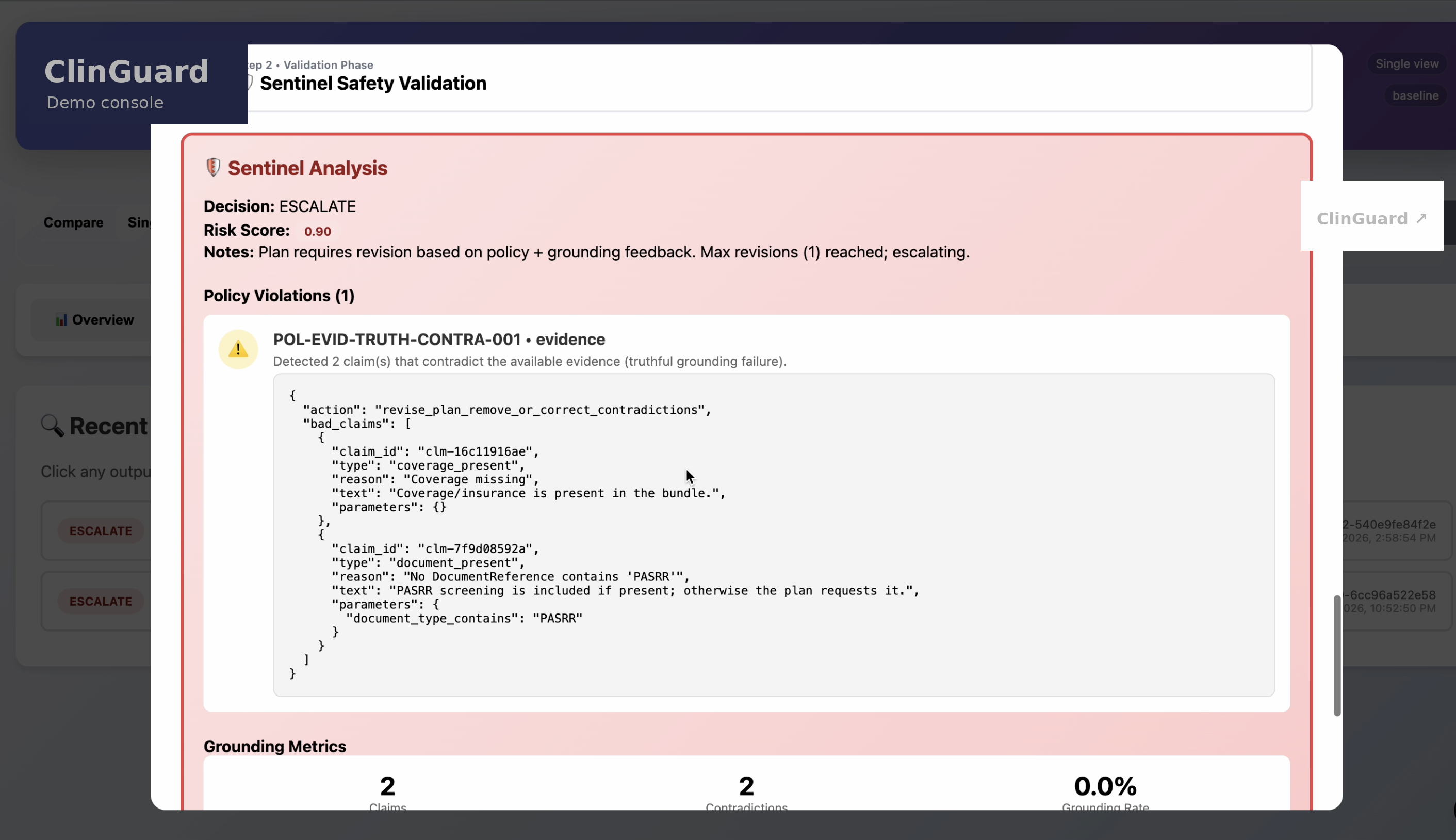}
\caption{\textbf{Sentinel safety validation.} The critic's decision (for example, escalate) with policy-violation details and the grounded sources used to justify it.}
\label{fig:sup-console-sentinel}
\end{figure}

\begin{figure}[p]
\centering
\includegraphics[width=0.98\linewidth]{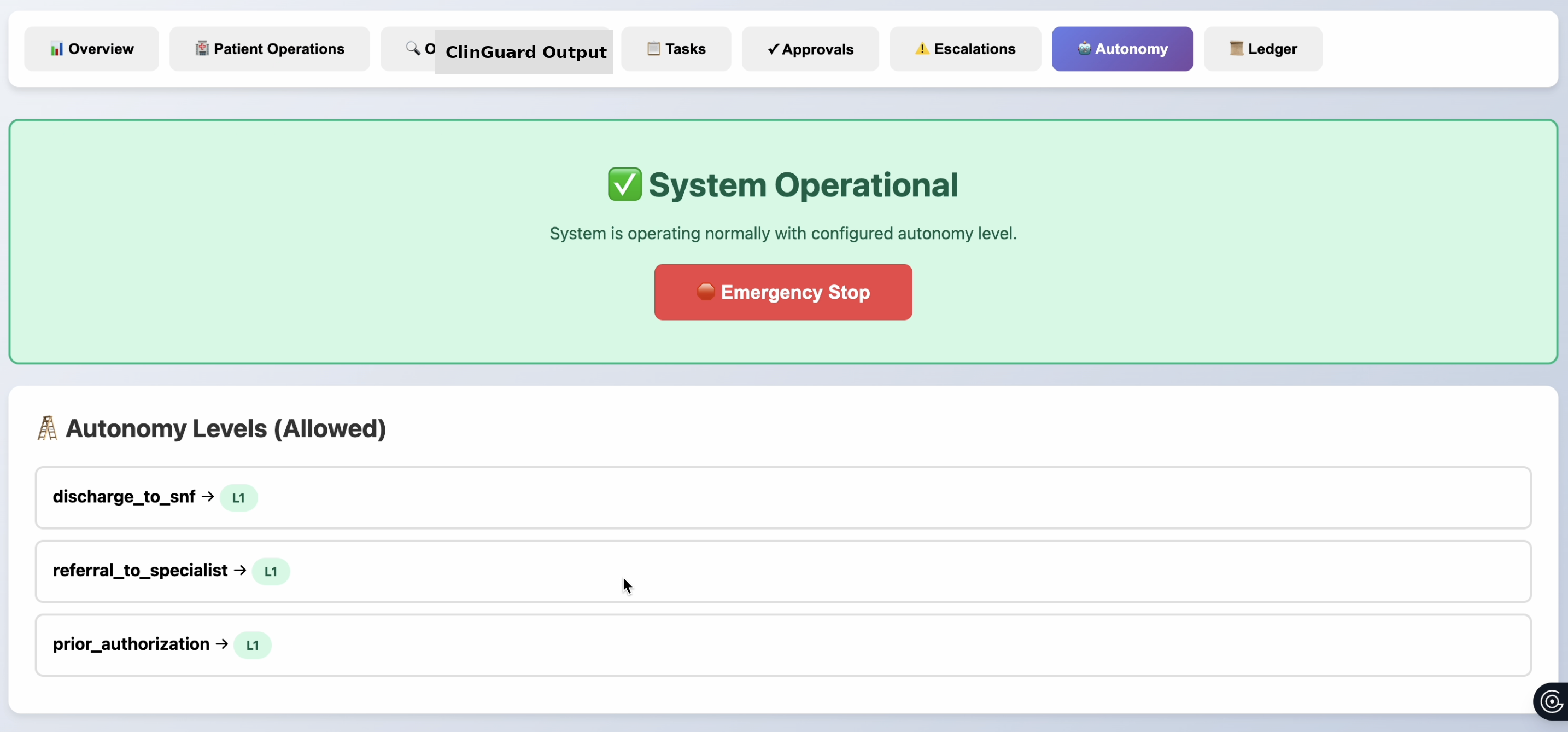}
\caption{\textbf{Governance control: autonomy and the circuit breaker.} Autonomy configuration and emergency stop, with ledger-integrity verification and audit entries available for trace-level forensics.}
\label{fig:sup-console-governance}
\end{figure}

The operator console is a thin user-interface layer over the transaction engine. It surfaces (i) patient-scoped outputs, (ii) task queues, (iii) safety validation and escalation decisions, (iv) autonomy controls (including an emergency stop), and (v) ledger-integrity verification and audit entries (Figs.~\ref{fig:sup-console-lists}, \ref{fig:sup-console-action-plan}, \ref{fig:sup-console-sentinel}, and \ref{fig:sup-console-governance}).

\clearpage

\section*{Appendix G. Sentinel surveillance checklist for Responsible Use of AI in Healthcare}
\label{sup:appendix-h-ruaih-mapping}

\begin{quote}\textit{Referenced in main manuscript Section on alignment with regulatory guidance.}\end{quote}

The Joint Commission and the Coalition for Health AI describe seven elements for the Responsible Use of AI in Healthcare. In practice, many of these elements are operational: they require a surveillance system that can (i) measure deployment behavior, (ii) detect drift and bias, (iii) capture safety events, and (iv) trigger governance action. This appendix provides an illustrative mapping (Table~\ref{tab:sup-ruaih-mapping}) between the seven elements and the concrete control points implemented by ClinGuard.

The mapping below complements the broader regulatory-alignment summary in Section 5.3 of the main paper, which positions ClinGuard against U.S.\ Food and Drug Administration, National Institute of Standards and Technology, and World Health Organization guidance for clinical artificial intelligence.

\begin{table}[t]
\centering
\caption{Mapping of Responsible Use of AI in Healthcare elements to ClinGuard control points and artifacts.}
\label{tab:sup-ruaih-mapping}
\scriptsize
\setlength{\tabcolsep}{3.5pt}
\renewcommand{\arraystretch}{1.1}
\begin{tabular}{p{0.23\linewidth} p{0.33\linewidth} p{0.38\linewidth}}
\hline
\textbf{Element} & \textbf{Sentinel surveillance and governance control point} & \textbf{Concrete artifact in this paper} \\
\hline
Artificial-intelligence policies and governance structures & Versioned, testable governance artifacts with named owners; dashboards summarizing outcomes events. & Policy registry (\texttt{Policy\_ID}s), pathway specifications, circuit-breaker thresholds, decision record and audit ledger. \\
\hline
Patient privacy and transparency & Minimum-necessary cross-tier contracts; access logging; pathway-specific transparency triggers (disclosure or consent). & Three-tier message contracts; minimum-necessary policy in the evidence adapter; access logs and ledger traces. \\
\hline
Data security and data-use protections & Encryption, strict access controls, connector allowlists, and audited data flows; deploy-time prerequisites. & Limited-privilege connectors; audit-ledger integrity envelope; governance-owned data-use constraints. \\
\hline
Ongoing quality monitoring & Continuous monitoring of latency, closure, safety intercepts, drift, and equity slices; governance triggers for sampling, downgrade, or kill switch. & Outcome specifications; outcomes-monitor dashboards; consensus entropy; edge-riding indicators. \\
\hline
Voluntary, blinded reporting of artificial-intelligence safety events & Safety-event bundling from intercepts and circuit-breaker activations; de-identification and optional external reporting. & Safety-intercept reason codes; circuit-breaker events; append-only hashes for de-identified event sharing. \\
\hline
Risk and bias assessment & Prospective documentation (limitations, intended use) plus post-deployment bias surveillance and local validation. & Model-card references; equity stratifiers in outcome specifications; subgroup monitoring and audits. \\
\hline
Education and training & Role-specific training and shared vocabulary embedded in workflows; teachable audit artifacts for review. & Operator console; decision-record summaries; governance playbooks for escalation and change control. \\
\hline
\end{tabular}
\end{table}

\subsection*{G1. Safety-event reporting bundle}
\label{supsub:h1-safety-event-reporting-bundle}

When the sentinel intercepts a proposal at high severity (VETO or ESCALATE) or activates a circuit breaker, the system can emit a safety-event report (a \texttt{SafetyEventReport}) suitable for incident review and (after de-identification) optional external reporting. The key requirement is that the report links to immutable hashes and governance versions rather than raw patient narrative.

\begin{codeblock}
SafetyEventReport = {
  "event_id": "SEV-UUID",
  "event_type": "VETO|ESCALATE|CIRCUIT_BREAKER|NEAR_MISS",
  "severity": "low|moderate|high|critical",
  "created_at_utc": "...",

  # Traceability without PHI leakage
  "trace_id": "...",
  "transaction_ids": ["..."],
  "pathway_id": "...",
  "pathway_version": "...",
  "policy_ids": [{"policy_id":"...","version":"..."}],
  "context_hash": "sha256:...",          # frozen snapshot pointer
  "proposal_hash": "sha256:...",
  "critique_hash": "sha256:...",
  "action_pack_hash": "sha256:...",

  # Human-reviewable summary (PHI-minimized)
  "summary": {
    "what_happened": "...",
    "why_intercepted": ["REASON_CODE_1", "REASON_CODE_2"],
    "recommended_human_action": "..."
  },

  # Optional: de-identified attachments for learning
  "attachments": {
    "decision_dossier_ref": "pointer_or_id",
    "redacted_evidence_refs": ["FHIR:Observation/..", "SOP:.."],
    "consensus_entropy_window": {"window":"7d","value":0.12}
  },

  "internal_reporting": {"system":"incident_reporting", "ticket_id":"..."},
  "external_reporting": {"enabled": false, "channel": "PSO_or_registry", "report_id": null}
}
\end{codeblock}

This schema is intentionally compatible with the paper's central claims: (i) the sentinel's gate decisions are deterministic conditional on versioned governance artifacts, (ii) privacy is protected through minimum-necessary data sharing, and (iii) post-deployment surveillance is audit-grade because it is tied to immutable hashes and explicit versions.

\clearpage

\section*{Appendix H. Full audit-ledger and decision-record schemas}
\label{sup:appendix-i-full-schemas}

\begin{quote}\textit{The conceptual overview of the audit-ledger envelope and the decision-record payload appears in the main paper's audit-artifacts table, and the integrity model those artifacts rest on is set out in the main paper's subsection on the ledger integrity model. The concrete field-level schemas are below (Tables~\ref{tab:sup-audit-ledger-full} and~\ref{tab:sup-decision-record-full}).}\end{quote}

\subsection*{H1. Audit-ledger envelope schema (full)}
\label{supsub:i1-audit-ledger-envelope-full}

Table~\ref{tab:sup-audit-ledger-full} gives the envelope written for every ledger entry. Its
fields fall into three groups: those that fix the entry's position in the hash chain, those that
bind it to the transaction and to the policy versions in force, and those that record what
produced the verdict.

{\scriptsize
\begin{longtable}{p{0.29\linewidth} p{0.35\linewidth} p{0.30\linewidth}}
\caption{Full audit-ledger envelope schema (append-only integrity and pointers).}\label{tab:sup-audit-ledger-full}\\
\hline
\textbf{Field} & \textbf{Description} & \textbf{Purpose} \\
\hline
\endfirsthead
\multicolumn{3}{l}{\footnotesize\textit{Audit-ledger envelope schema, continued from the previous page.}}\\
\hline
\textbf{Field} & \textbf{Description} & \textbf{Purpose} \\
\hline
\endhead
\hline
\multicolumn{3}{r}{\footnotesize\textit{continued on the next page}}\\
\endfoot
\hline
\endlastfoot
\texttt{trace\_id} & UUIDv4 for the coordination episode (spans from trigger to closure). & End-to-end grouping for an episode. \\
\hline
\texttt{transaction\_id} & UUIDv4 for a single state-transition attempt. & Uniquely identifies revisions and retries. \\
\hline
\texttt{context\_hash} & Hash of the context information retrieved (data plus timestamps). & Reproducibility; prevents moving-target ambiguity. \\
\hline
\texttt{proposal\_hash} & Hash of the coordinator agent's structured proposal payload. & Anti-tamper proof of what was proposed. \\
\hline
\texttt{critique\_hash} & Hash of the sentinel's structured critique and verdict payload. & Anti-tamper proof of what was validated. \\
\hline
\texttt{action\_pack\_hash} & Hash of compiled action contents. & Binds authorization to concrete executable actions. \\
\hline
\texttt{policy\_ids} & List of policy identifiers and versions applied. & Traceability to governance artifacts. \\
\hline
\texttt{pathway\_id\_version} & Pathway-specification identifier and version. & Ensures pathway versioning at execution time. \\
\hline
\texttt{decision\_dossier\_ref} & Pointer or identifier to the stored decision record. & Links envelope to semantic payload. \\
\hline
\texttt{sentinel\_verdict} & APPROVE / REVISE / VETO / ESCALATE plus reason codes. & Accountability for authorization. \\
\hline
\texttt{timestamp\_utc} & UTC timestamp when the envelope was appended. & Latency and quality-of-service monitoring; audit timeline. \\
\hline
\texttt{prev\_ledger\_hash} & Hash pointer to the prior envelope. Required, not optional: the chain is the integrity mechanism, so an envelope written without it cannot be verified. & Append-only integrity and tamper evidence. \\
\hline
\texttt{envelope\_signature} & Digital signature over the envelope contents, applied by the assurance tier at commit time. & Attributes the entry to the component that wrote it; detects substitution of a whole envelope. \\
\hline
\texttt{signing\_key\_id} & Identifier and version of the key used to sign. & Permits key rotation without invalidating earlier entries. \\
\hline
\texttt{anchor\_ref} & Identifier of the most recent chain-head digest published to an external witness, and its publication time. & Makes truncation of the chain's tail detectable by a party outside the operating team. \\
\hline
\texttt{evaluator\_image\_digest} & Digest of the container image of the sentinel build that produced the verdict. & Lets an auditor reconstruct a decision with the exact evaluator that made it, not a later build. \\
\end{longtable}
}

\subsection*{H2. Decision-record payload schema (full)}
\label{supsub:i2-decision-record-payload-full}

Table~\ref{tab:sup-decision-record-full} gives the payload each entry carries. Where the envelope
in Table~\ref{tab:sup-audit-ledger-full} establishes that an entry is authentic and correctly
ordered, the payload is what an auditor reads: the context the agent saw, the proposal it made,
the critique that judged it, and the action that followed.

{\small
\begin{longtable}{p{0.22\linewidth} p{0.42\linewidth} p{0.30\linewidth}}
\caption{Full decision-record payload schema.}\label{tab:sup-decision-record-full}\\
\hline
\textbf{Section} & \textbf{Key fields (examples)} & \textbf{Why it matters} \\
\hline
\endfirsthead
\multicolumn{3}{l}{\footnotesize\textit{Decision-record payload schema, continued from the previous page.}}\\
\hline
\textbf{Section} & \textbf{Key fields (examples)} & \textbf{Why it matters} \\
\hline
\endhead
\hline
\multicolumn{3}{r}{\footnotesize\textit{continued on the next page}}\\
\endfoot
\hline
\endlastfoot
Identifiers & \texttt{trace\_id}, \texttt{transaction\_id}, \texttt{pathway\_id}, \texttt{pathway\_version}, \texttt{task\_type} & Unambiguous linkage between episode, transaction, and governance artifacts. \\
\hline
Trigger and context (inputs and provenance) & \texttt{trigger\_event}, \texttt{context\_timestamp\_utc}, \texttt{fhir\_pointers}, \texttt{context\_hash} & Reconstructs what the system saw at the moment of decision. \\
\hline
Evidence bundle & \texttt{evidence\_items[]} (source, identifier, strength, applicability, contraindications, key findings) & Makes evidence provenance explicit; supports citation and adjudication. \\
\hline
Rules and pathway clauses & \texttt{policies\_applied[]}, \texttt{pathway\_clauses\_triggered[]} (with versions) & Operationalizes compliance-by-construction and enables governance updates. \\
\hline
Tradeoffs and alternatives & \texttt{options\_considered[]}, \texttt{conflicts[]}, \texttt{rejected\_reason\_codes} & Captures `why A not B' in structured form; supports audit and learning. \\
\hline
Risk tier and sign-offs (confirmation chain) & \texttt{risk\_tier}, \texttt{required\_signoffs[]}, \texttt{autonomy\_level} & Aligns with clinical expectations and liability boundaries. \\
\hline
Authorization outcome & \texttt{sentinel\_verdict}, \texttt{revise\_directives[]} (if any) & Documents APPROVE, REVISE, and VETO decisions and required remediation. \\
\hline
Action and execution & \texttt{action\_pack[]}, \texttt{write\_targets[]}, \texttt{execution\_status}, \texttt{rollback\_events[]} & Binds recommendation to concrete actions and their outcomes. \\
\hline
Monitoring hooks & \texttt{outcome\_spec\_id}, \texttt{slo\_targets}, \texttt{equity\_stratifiers}, \texttt{consensus\_rate}, \texttt{consensus\_entropy} & Enables time-stamped evaluation, subgroup monitoring, and auditing for abnormal agreement patterns (mode collapse or gaming). \\
\hline
Append-only integrity & \texttt{dossier\_hash}, \texttt{prev\_ledger\_hash}, \texttt{ledger\_timestamp\_utc} & Tamper-evident logging and forensic traceability. \\
\end{longtable}
}

The conceptual overviews of these two schemas appear in the main paper to keep prose-level reading flow clear of identifier-level detail; the full identifier-level field schemas above are intended for downstream implementers and auditors who need to inspect the data contracts.

\clearpage

\section*{Appendix I. Worked-example timeline (one skilled-nursing-facility transaction)}
\label{sup:appendix-j-worked-example-timeline}

\begin{quote}\textit{Referenced in main manuscript Section on the worked example.}\end{quote}

\begin{figure}[htbp]
\centering
\includegraphics[width=\linewidth]{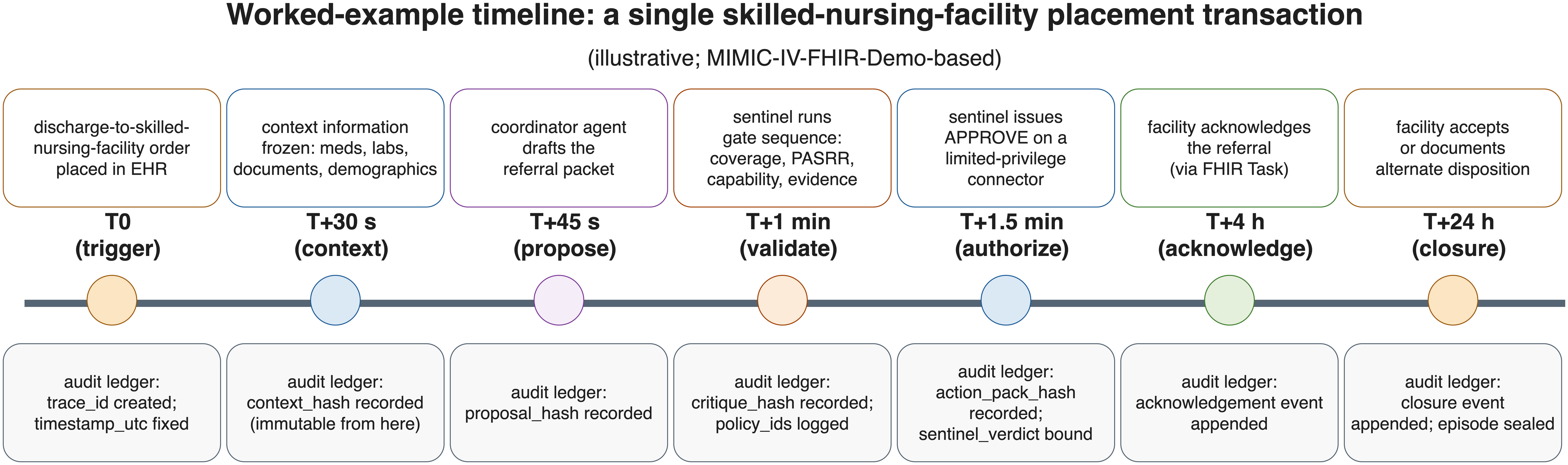}
\caption{\textbf{Worked-example timeline: a single skilled-nursing-facility placement transaction.} Above the spine: workflow events at each waypoint. Below the spine: corresponding append-only audit-ledger records, hash-bound and immutable from the moment of capture. Timings are illustrative and based on the MIMIC-IV-FHIR dataset.}
\label{fig:sup-timeline}
\end{figure}

Figure~\ref{fig:sup-timeline} traces a single skilled-nursing-facility placement transaction minute by minute. Above the spine: what happens in the workflow at each waypoint. Below the spine: the immutable record the audit ledger captures at that waypoint. The same shape applies, with adjusted timing and acuity, to other coordination workflows in the family (hospital-to-home transitions, prior authorization, specialty referral, transitional heart-failure, rural-health hand-offs).

\clearpage

\section*{Appendix J. Per-workflow evaluation metrics for the six coordination scenarios}
\label{sup:appendix-j-per-workflow-evaluation-metrics}

\begin{quote}\textit{Referenced in main manuscript Section~6.1 (``Why the family of coordination workflows is the right wedge'').}\end{quote}

Table~\ref{tab:sup-clinical-scenarios} maps each coordination workflow to its principal failure mode, sentinel control, candidate key performance indicators (KPIs), default risk tier, and primary sentinel pillar. These are evaluation targets, not reported outcomes. Only skilled-nursing-facility placement is implemented with MIMIC-IV-FHIR data and included in the planned Duke stepped-wedge trial; the other rows describe prospective extensions.

\begin{landscape}
\begin{table}[p]
\centering
\caption{Per-workflow evaluation roadmap: candidate metrics for the six coordination scenarios.}
\label{tab:sup-clinical-scenarios}
\small
\setlength{\tabcolsep}{3pt}
\renewcommand{\arraystretch}{0.89}
\begin{tabular}{p{0.16\linewidth} p{0.17\linewidth} p{0.25\linewidth} p{0.20\linewidth} p{0.07\linewidth} p{0.07\linewidth}}
\hline
\textbf{Scenario} & \textbf{Failure mode today} & \textbf{Sentinel mechanism on the critical path} & \textbf{Candidate KPIs} & \textbf{Tier} & \textbf{Pillar} \\
\hline
Skilled-nursing-facility placement \newline (Inpatient $\to$ SNF) &
Manual chart mining; fragmented portals; payer--capability mismatch. &
$\bullet$ Coordinator compiles and tracks state. \newline
$\bullet$ Sentinel enforces coverage, pre-admission-screening, and capability gates. &
$\bullet$ Time to acceptance. \newline
$\bullet$ Rework rate. \newline
$\bullet$ Avoidable length of stay. \newline
$\bullet$ Equity by group. &
Medium & Govern \\
\hline
Hospital-to-home transition \newline (Inpatient $\to$ Home) &
Medication discrepancies, missed follow-up, weak teach-back. &
$\bullet$ Coordinator drafts the bundle and follow-ups. \newline
$\bullet$ Sentinel audits high-risk medications and teach-back. &
$\bullet$ Packet completion time. \newline
$\bullet$ Discrepancy rate. \newline
$\bullet$ Seven-day follow-up rate. &
Medium & Monitor \\
\hline
Prior authorization and medication refill \newline (Order $\to$ Payer approval) &
Documentation burden; denial-and-appeal cycles. &
$\bullet$ Coordinator drafts to payer criteria. \newline
$\bullet$ Sentinel cross-checks evidence and submission latency. &
$\bullet$ Physician hours on prior authorization. \newline
$\bullet$ First-pass approval rate. \newline
$\bullet$ Time to decision. &
Low \newline to high\textsuperscript{*} & Govern \\
\hline
Specialty referral closure \newline (Primary care $\to$ Specialist) &
Lost referrals; scheduling delays; lost-to-follow-up. &
$\bullet$ Coordinator logs and attaches data. \newline
$\bullet$ Sentinel monitors acknowledgement latency. &
$\bullet$ Time to acknowledgement ($<2$ d). \newline
$\bullet$ Completion rate. \newline
$\bullet$ Patient satisfaction. &
Low & Validate \\
\hline
Transitional heart-failure care \newline (Admission $\to$ 30 d) &
Missed guideline-directed therapy; delayed follow-up; fill gaps. &
$\bullet$ Coordinator instantiates the heart-failure checklist. \newline
$\bullet$ Sentinel validates contraindications and closure. &
$\bullet$ 30-day readmission. \newline
$\bullet$ Discharge on guideline-directed therapy. \newline
$\bullet$ Follow-up plus fill latency. &
High & Monitor \\
\hline
Rural-health hand-off \newline (Rural $\to$ Tertiary) &
Distance plus vendor heterogeneity; acknowledgement gaps. &
$\bullet$ Coordinator compiles the cross-vendor packet. \newline
$\bullet$ Sentinel verifies acceptance and escalates. &
$\bullet$ Time to tertiary-site acceptance. \newline
$\bullet$ Documentation completeness on receipt. \newline
$\bullet$ Avoided redundant transfers. &
High & Escalate \\
\hline
\end{tabular}

\smallskip
{\footnotesize \emph{*} Prior-authorization risk tier varies with the request: routine renewals and benign medication refills sit at the low end, while high-acuity authorizations (chemotherapy, biologics, prior-auth-gated procedures) sit at the high end.}
\end{table}
\end{landscape}

Every KPI must be reconstructable from audit-ledger timestamps. Analyses should be stratified by language, insurance, site, unit, and protected demographic groups where available. The default risk tier initializes autonomy settings; pathway-defined acuity triggers may reclassify an individual transaction, as with prior authorization for a high-acuity chemotherapy regimen.

\clearpage

\section*{Appendix K. How thresholds, predicates, and escalation conditions are set and checked}
\label{sup:appendix-k-parameter-determination}

\begin{quote}\textit{Referenced in main manuscript Section on the sentinel critic implementation.}\end{quote}

This appendix specifies how thresholds are calibrated, who may change them, which tests predicates must pass, and what the registry records for later audit.

\noindent{\large\bfseries K1. Two classes of parameter}\par\medskip
\label{supsub:k1-two-classes-of-parameter}

Every configurable quantity is either a deployment-wide invariant or a locally adjustable tunable constrained to a declared, versioned range.

\begin{table}[htbp]
\centering
\caption{Governance invariants and tunable parameters.}
\label{tab:sup-invariants}
\small
\setlength{\tabcolsep}{3pt}
\begin{tabular}{p{0.15\linewidth} p{0.40\linewidth} p{0.36\linewidth}}
\hline
\textbf{Class} & \textbf{What it covers} & \textbf{Change control} \\
\hline
Invariant &
$\bullet$ The gate sequence runs before every state-changing action. \newline
$\bullet$ No claim is accepted without a source pointer. \newline
$\bullet$ An unresolved uncertainty never resolves to approval. \newline
$\bullet$ Every verdict is written to the ledger. \newline
$\bullet$ Every pathway names an accountable owner and an escalation route. &
No deployment role may change an invariant. Framework-level changes require institutional governance approval, and the manifest registry rejects any specification that omits one. \\
\hline
Tunable, within a declared range &
$\bullet$ Confidence cutoffs for concept resolvers. \newline
$\bullet$ Latency targets and timeout windows. \newline
$\bullet$ The risk tier assigned to a task type. \newline
$\bullet$ Which role reviews which class of exception. \newline
$\bullet$ Monitoring window lengths and deviation limits. &
The pathway owner may change a tunable with safety and operations board approval. The registry rejects out-of-range values, and each change must pass the K3 regression corpus. \\
\hline
\end{tabular}
\end{table}

\subsection*{K2. Calibrating a threshold}
\label{supsub:k2-calibrating-a-threshold}

The same procedure applies to any tunable cutoff; a concept-resolver confidence threshold is used as the example.

\begin{enumerate}
\item \textbf{Start in shadow mode.} Use the conservative default (0.9 in this example), record every verdict, and execute no action.
\item \textbf{Accumulate a comparison set.} Collect enough transactions to estimate false escalation and missed interception; a few hundred per pathway and unit is the working minimum, not a validated universal sample size.
\item \textbf{Obtain independent adjudication.} Clinicians who did not author the pathway review a stratified sample enriched near the cutoff. Their judgments form the reference standard, and disagreement is retained as evidence of under-specification.
\item \textbf{Select and document an operating point.} The pathway owner balances alert burden against misses. Higher-risk or irreversible actions use tighter cutoffs; the selected value and rationale become part of the governed artifact.
\item \textbf{Promote and monitor.} Level 1 promotion requires review. If an alert-fatigue indicator crosses its ceiling, the threshold returns to shadow mode and requires new adjudication before re-promotion.
\item \textbf{Recalibrate after material change.} A new pathway, policy set, or acting model invalidates the baseline.
\end{enumerate}

\subsection*{K3. What a policy predicate must pass before it can block anything}
\label{supsub:k3-predicate-acceptance}

The manifest registry requires three test classes before a predicate can be promoted:

\begin{itemize}
\item \textbf{Must-fire cases:} transactions the predicate must catch.
\item \textbf{Must-not-fire cases:} legitimate transactions that must pass.
\item \textbf{Must-abstain cases:} transactions whose inputs do not resolve the predicate.
\end{itemize}

Every change runs these cases and a stored transaction corpus. Newly blocked transactions require review; newly allowed transactions require named approval before promotion.

\subsection*{K4. Uncertainty and escalation}
\label{supsub:k4-uncertainty-and-escalation}

A resolver returns holds, does not hold, or unresolved. Unresolved results route by risk tier: administrative ambiguity to a navigator and high-risk clinical ambiguity to a credentialed clinician. The ledger records both the resolver output and human decision, adding adjudicated cases to the calibration set.

\subsection*{K5. What is recorded, and why it matters later}
\label{supsub:k5-what-is-recorded}

For each tunable, the registry retains its value and range, applicable pathway and unit, adjudicated sample, operating-point rationale, approver, and latest recalibration date. These fields support reconstruction of both the setting and its justification.

\section*{Appendix L. Supporting bench-scale results}
\label{supsec:appendix-l-bench-evaluation}

This appendix reports supporting results for the evaluation design in the main text. Captions identify each result source.

\subsection*{L1. Per-gate latency decomposition}
\label{supsub:l1-gate-latency}

At peak measured concurrency, the model call was approximately three orders of magnitude slower than any deterministic gate (Table~\ref{tab:sup-gate-latency}).

\begin{table}[htbp]
\centering
\caption{Per-gate latency decomposition of a governed transaction at concurrency 32. Share is of total per-transaction gate time. Computed from \texttt{phase2\_summary.json}.}
\label{tab:sup-gate-latency}
\small
\begin{tabular}{lrrr}
\hline
\textbf{Gate} & \textbf{Median (ms)} & \textbf{p95 (ms)} & \textbf{Share} \\
\hline
Plan proposal (model call) & 26813.76 & 43972.90 & 99.703\% \\
Context assembly & 51.71 & 427.54 & 0.192\% \\
Sentinel critique & 12.96 & 110.28 & 0.048\% \\
Ledger append & 9.09 & 140.30 & 0.034\% \\
Autonomy resolution & 4.61 & 8.00 & 0.017\% \\
Task initialization & 1.05 & 62.73 & 0.004\% \\
Evidence retrieval & 0.43 & 0.49 & 0.002\% \\
Escalation drafting & 0.13 & 0.28 & 0.000\% \\
\hline
\end{tabular}
\end{table}

\subsection*{L2. Diurnal surge arm}
\label{supsub:l2-surge}

The surge arm compressed the observed hour-of-day distribution of 275 encounter starts into the bench window, with a 1.8$\times$ multiplier applied at the 07:00, 15:00, and 23:00 nursing shift boundaries, so the profile peaks at those hours.

\begin{table}[htbp]
\centering
\caption{Diurnal surge arm. Computed from \texttt{surge\_summary.json}.}
\label{tab:sup-surge}
\small
\begin{tabular}{lr}
\hline
\textbf{Measure} & \textbf{Value} \\
\hline
Cycle window & 720 s \\
Peak concurrent users & 24 \\
Transactions completed & 198 \\
Failed transactions & 0 \\
Throughput & 16.47 min$^{-1}$ \\
End-to-end latency, median & 25.3 s \\
End-to-end latency, p95 & 42.2 s \\
End-to-end latency, p99 & 50.9 s \\
Governance latency, median & 92.9 ms \\
Governance share of transaction & 4.03\% \\
Encounters shaping the profile & 275 \\
Busiest hours in the profile & 23:00, 07:00, 15:00 \\
\hline
\end{tabular}
\end{table}

\subsection*{L3. Ledger scaling, chain verification, and breaker actuation}
\label{supsub:l3-ledger}

Append latency remained stable with chain depth because the tail digest is retained in memory and revalidated against file size; an externally altered file is reread before append.

\begin{table}[htbp]
\centering
\caption{Ledger append cost as a function of chain depth, with chain verification and autonomy-breaker actuation. Computed from \texttt{phase0\_instrumentation.json}.}
\label{tab:sup-ledger-scaling}
\small
\begin{tabular}{lrrr}
\hline
\textbf{Chain depth (entries)} & \textbf{Append p50 (ms)} & \textbf{Append p95 (ms)} & \textbf{Appends s$^{-1}$} \\
\hline
100 & 0.034 & 0.047 & 29{,}197 \\
1{,}000 & 0.032 & 0.045 & 31{,}048 \\
10{,}000 & 0.038 & 0.051 & 26{,}172 \\
50{,}000 & 0.038 & 0.040 & 26{,}622 \\
100{,}000 & 0.039 & 0.050 & 25{,}504 \\
\hline
\end{tabular}
\smallskip

{\footnotesize \textbf{Chain verification:} 100,000 entries (48.4 MB) in 656 ms; 152,384 entries s$^{-1}$; integrity verified. \textbf{Autonomy actuation:} kill switch, 0.131 ms; veto-rate breaker, 0.472 ms after its triggering condition.}
\end{table}

\subsection*{L4. Transaction-conformance fault suite}
\label{supsub:l4-fault-suite}

Five mutations were injected: missing prerequisite, fabricated policy identifier, prohibited transition, stale context hash, and silent partial completion. Interception required the expected check to fire on the mutant but not its paired control.

Table~\ref{tab:sup-fault-interception} reports paired interception and verdict escalation. Missing-prerequisite interception is lower because neither corpus contains the required preadmission documentation, so the control already triggers the prerequisite policy. The verdict-escalation column shows whether the mutation nevertheless changed the decision from revision to hard block.

Each mutation class corresponds to a failure mode described in the care-coordination or language-model literature. The classes are not arbitrary perturbations. Table~\ref{tab:sup-fault-mapping} summarizes these relationships, with supporting citations provided in Section~7 of the main text. Because the classes were derived from the gate contract, they systematically exercise the framework's checks but do not capture every failure that may occur in practice. An independently derived fault set will instead be developed from institutional incident reports and clinician-labeled faults collected during the planned shadow phase.

\begin{table}[htbp]
\centering
\caption{Correspondence between injected fault classes and documented coordination failure modes. Bracketed numbers refer to the reference list of the main text.}
\label{tab:sup-fault-mapping}
\small
\setlength{\tabcolsep}{4pt}
\begin{tabular}{p{0.22\linewidth} p{0.36\linewidth} p{0.34\linewidth}}
\hline
\textbf{Fault class} & \textbf{Mutator operationalization} & \textbf{Documented failure mode} \\
\hline
Missing prerequisite & removes prerequisite content the proposal asserted (for example, the preadmission-screening assertion) & transfers and discharges that proceed without required documentation [9, 51] \\
Fabricated policy identifier & adds a citation to a policy and guideline that do not exist & model-fabricated guideline, order-set, or policy identifiers [11, 12] \\
Prohibited transition & advances the task graph across an unmet checkpoint & workflow steps executed before their preconditions are met \\
Stale context hash & presents the proposal under a superseded context snapshot & actions based on patient state that has since changed [8, 52] \\
Silent partial completion & reports a multi-step action bundle as complete after removing steps & referral and follow-up loops reported closed without the underlying work completed [25, 26] \\
\hline
\end{tabular}
\end{table}

\begin{table}[htbp]
\centering
\caption{Interception of injected transaction non-conformance, both corpora. Intervals are Wilson 95\% intervals. Computed from \texttt{phase3\_summary.json} and \texttt{fault\_rows.jsonl} of each arm.}
\label{tab:sup-fault-interception}
\small
\setlength{\tabcolsep}{4pt}
\begin{tabular}{lrrlr}
\hline
\textbf{Fault class} & \textbf{N} & \textbf{Intercepted} & \textbf{95\% interval} & \textbf{Verdict escalated} \\
\hline
\multicolumn{5}{l}{\emph{Synthea-derived corpus, 200 frozen proposals}} \\
Missing prerequisite & 200 & 56.5\% & 49.6--63.2\% & 95\% \\
Fabricated policy identifier & 200 & 99.0\% & 96.4--99.7\% & 0\% \\
Prohibited transition & 200 & 100.0\% & 98.1--100\% & 0\% \\
Stale context hash & 200 & 100.0\% & 98.1--100\% & 0\% \\
Silent partial completion & 200 & 100.0\% & 98.1--100\% & 0\% \\
\textbf{Pooled} & \textbf{1000} & \textbf{91.1\%} & \textbf{89.2--92.7\%} & \\
\hline
\multicolumn{5}{l}{\emph{MIMIC-IV-FHIR demonstration corpus, 100 frozen proposals}} \\
Missing prerequisite & 100 & 37.0\% & 28.2--46.8\% & 91\% \\
Fabricated policy identifier & 100 & 100.0\% & 96.3--100\% & 0\% \\
Prohibited transition & 100 & 100.0\% & 96.3--100\% & 0\% \\
Stale context hash & 100 & 100.0\% & 96.3--100\% & 0\% \\
Silent partial completion & 100 & 100.0\% & 96.3--100\% & 0\% \\
\textbf{Pooled} & \textbf{500} & \textbf{87.4\%} & \textbf{84.2--90.0\%} & \\
\hline
\end{tabular}
\end{table}

In the Synthea control arm, three checks never fired and two fabricated-identifier alerts were verified as malformed identifiers generated by the planner (Table~\ref{tab:sup-specificity}). Missing-prerequisite specificity is not estimable from this corpus because every control lacks the required document; the displayed 0\% is therefore a corpus property, not a deployment false-positive estimate.

\begin{table}[htbp]
\centering
\caption{Specificity of each conformance check on unmutated proposals, Synthea-derived arm. Computed from the control arm of \texttt{fault\_rows.jsonl}.}
\label{tab:sup-specificity}
\small
\begin{tabular}{lrrr}
\hline
\textbf{Check} & \textbf{Control proposals} & \textbf{Fired} & \textbf{Specificity} \\
\hline
Prohibited transition & 200 & 0 & 100.0\% \\
Stale context hash & 200 & 0 & 100.0\% \\
Silent partial completion & 200 & 0 & 100.0\% \\
Fabricated policy identifier & 200 & 2 & 99.0\% \\
Missing prerequisite & 200 & 200 & not estimable$^{*}$ \\
\hline
\end{tabular}
\smallskip

{\footnotesize $^{*}$All controls lacked the required document; the observed pass rate was 0\%, so deployment specificity cannot be estimated from this corpus.}
\end{table}

\subsection*{L5. Safeguarding contribution and sub-gate attribution}
\label{supsub:l5-safeguarding}

Table~\ref{tab:sup-safeguarding} reports supervisory outcomes across 674 fully instrumented concurrency-sweep transactions. Revision-loop yield counts violations resolved between successive critiques of the same transaction.

The sign-off sub-gate was not exercised because these proposals had no required-signoff entries. Medium-severity grounding findings dominated escalation, making the reported precision a calibration result rather than a safety estimate. Because no transaction was approved, approval-rate disparities are not estimable.

\begin{table}[htbp]
\centering
\caption{Safeguarding contribution of the supervisory layer over 674 governed transactions. Computed from the per-transaction records and trust ledger of the concurrency sweep.}
\label{tab:sup-safeguarding}
\small
\begin{tabular}{lr}
\hline
\textbf{Measure} & \textbf{Value} \\
\hline
\multicolumn{2}{l}{\emph{Revision-loop yield}} \\
Violations resolved by revision & 1361 \\
Violations resolved per revised transaction, mean & 2.08 \\
Revised transactions with positive yield & 94.4\% \\
Transactions entering the revision loop & 656 \\
Revision rounds per transaction, mean & 2.53 \\
\hline
\multicolumn{2}{l}{\emph{Attribution by critique sub-gate}} \\
Symbolic policy rules & 936 violations (47.8\%) \\
Truthful-grounding engine & 1023 violations (52.2\%) \\
Guardrail sign-off validation & 0 violations (not exercised) \\
\hline
\multicolumn{2}{l}{\emph{Interception volume}} \\
Transactions evaluated & 674 \\
Violations per transaction, mean & 2.91 \\
Hard blocks per transaction, mean & 0.20 \\
Distinct policies exercised & 7 \\
Proposals routed to revise or veto before execution & 100.0\% \\
\hline
\multicolumn{2}{l}{\emph{Escalation and equity}} \\
Transactions escalated or vetoed & 674 \\
Escalations carrying at least one hard block & 20.5\% \\
Escalations with a safety or financial hard block & 5.0\% \\
Transactions approved for execution & 0 \\
Approval-rate disparity across slices & not estimable \\
\hline
\end{tabular}
\end{table}

\subsection*{L6. Model usage}
\label{supsub:l6-usage}

Table~\ref{tab:sup-usage} reports ledger-counted model calls and tokens. Calls per transaction exceed one because a revision directive can trigger a bounded re-proposal loop. The usage ledger contains 749 load-workload transaction records; Tables~\ref{tab:sup-safeguarding} and the main-text concurrency table report the 674 fully instrumented records retained for stage-level performance analysis. The difference reflects warm-up transactions issued before each measured window and transactions still in flight when a window closed, which the ledger records but the performance analyses exclude; the fault-suite ledgers likewise include warm-up calls beyond the 200 and 100 frozen proposal--control pairs analyzed.

\begin{table}[htbp]
\centering
\caption{Measured model usage across the evaluation. Cost is derived from measured tokens at a blended 2.32 USD per million tokens.}
\label{tab:sup-usage}
\small
\begin{tabular}{lrrrrr}
\hline
\textbf{Workload} & \textbf{Trans.} & \textbf{Calls} & \textbf{Calls/tx} & \textbf{Tokens} & \textbf{USD} \\
\hline
Load-workload ledger & 749 & 2654 & 3.54 & 12{,}086{,}969 & 28.04 \\
Fault suite, Synthea-derived & 201 & 792 & 3.94 & 4{,}774{,}712 & 11.08 \\
Fault suite, MIMIC-IV-FHIR & 100 & 375 & 3.75 & 1{,}801{,}965 & 4.18 \\
\hline
\textbf{Total} & & & & \textbf{18{,}663{,}646} & \textbf{43.30} \\
\hline
\end{tabular}
\end{table}

\clearpage
\noindent{\large\bfseries L7. What this evaluation does not establish}\par\medskip
\label{supsub:l7-boundaries}

The measurements reflect one workstation, one hosted model interface, and one model backbone. The five injected fault classes are not exhaustive, and their rates verify mechanisms rather than sensitivity to naturally occurring errors. No transaction reached approval, so the approval path and transaction-level specificity remain unmeasured. No patient was treated, clinical decision enacted, or clinical outcome measured.

\clearpage
\section*{Appendix M. Porting checklist for a new institution}
\label{sup:appendix-m}

The runtime itself does not need to be changed when the framework is deployed at a second institution. What must be evaluated locally are the available data, the policy configuration, the permitted execution surface, and the resulting performance baselines. The following sequence defines that adaptation process.

\begin{enumerate}
\item \textbf{Resource-availability audit.} Document the local FHIR version, licensed modules, populated resource types, code systems, and extensions. Then determine which policy prerequisites can be observed reliably from those resources. The benchmark corpora show why this step is necessary: neither includes preadmission-screening documentation, causing the corresponding prerequisite policy to fire even for every unmutated control (Appendix~L4).
\item \textbf{Policy-manifest re-parameterization.} Adapt pathway specifications, policy predicates, thresholds, sign-off roles, and escalation routes to the local setting. These elements should be maintained as versioned policy artifacts with clearly identified owners (Appendix~K). The sentinel code remains unchanged; each local adaptation is recorded as a new manifest version.
\item \textbf{Connector scoping.} Register the institution's execution surface using least-privilege write scopes, and verify that vendor write-back is available for every supported action type. Write-back capability must be confirmed during deployment rather than assumed by the architecture (main text, Limitations).
\item \textbf{Frozen-corpus replay.} Create a local retrospective corpus and freeze its proposal--evidence--context triples. Replay the fault suite and load profile against this corpus to establish local baselines for interception, specificity, and latency before granting execution authority. Model calls are made once when the corpus is frozen; subsequent replays are deterministic (Appendix~L4).
\item \textbf{Shadow-mode sign-off.} Run the system in Level~0 shadow mode on live data streams. The local governance committee should review the resulting decision records and escalations, with Level~1 authority granted only after the pre-specified criteria have been met (main text, Section~7).
\end{enumerate}

Steps 1--3 address local configuration and contracting requirements. Steps 4 and 5 repeat the gated validation sequence used in this study with data from the new institution. In this way, the portability assessment also serves as the site-specific validation process.

\end{document}